\documentclass[aps,prd,superscriptaddress,tighten,nofootinbib,twocolumn]{revtex4}

\usepackage{float}
\usepackage{amssymb}
\usepackage{amsmath,color}
\usepackage{graphicx}
\usepackage{bbm}
\usepackage{enumitem}

\begin{document}
\title{Status of the Zee--Babu model for neutrino mass and\\
possible tests at a like-sign linear collider}
\date{\today}
\author{Daniel Schmidt}
\email{danielschmidtip@t-online.de}

\affiliation{Max-Planck-Institut
f\"ur Kernphysik, Saupfercheckweg 1, 69117 Heidelberg, Germany}

\author{Thomas Schwetz}
\email{schwetz@fysik.su.se}

\affiliation{Max-Planck-Institut
f\"ur Kernphysik, Saupfercheckweg 1, 69117 Heidelberg, Germany}

\affiliation{Oskar Klein Centre for Cosmoparticle Physics,
Department of Physics, Stockholm University, SE-10691 Stockholm, Sweden}

\author{He Zhang}
\email{he.zhang@mpi-hd.mpg.de}

\affiliation{Max-Planck-Institut
f\"ur Kernphysik, Saupfercheckweg 1, 69117 Heidelberg, Germany}

\begin{abstract}
We provide an updated scan of the allowed parameter space of the
two-loop Zee--Babu model for neutrino mass. Taking into account most
recent experimental data on $\mu\to e\gamma$ as well as the mixing
angle $\theta_{13}$ we obtain lower bounds on the masses of the singly
and doubly charged scalars of between 1 to 2 TeV, with some dependence
on perturbativity and fine-tuning requirements. This makes the scalars
difficult to observe at LHC with 14~TeV even with optimistic
assumptions on the luminosity, and would require a multi-TeV linear
collider to see the scalar resonances. We point out, however, that a
sub-TeV linear collider in the like-sign mode may be able to observe
lepton flavour violating processes such as $e^- e^- \to \mu^- \mu^-$
due to contact interactions induced by the doubly charged scalar with
masses up to around 10~TeV. We investigate the possibility to
distinguish the Zee--Babu model from the Higgs triplet model using
such processes.
\end{abstract}

\maketitle

\section{Introduction}
\label{sec:introduction}

Non-zero neutrino mass requires an extension of the Standard
Model (SM). Among the plenitude of possibilities, an attractive way to
explain the smallness of neutrino masses is to invoke loop processes,
see for instance \cite{Farzan:2012ev, Angel:2012ug, Law:2013dya} for recent
discussions. Then the scale of the new physics responsible for
generating neutrino mass can be not too far from the TeV range, which
makes those type of models potentially testable at colliders and/or in
experiments searching for charged lepton flavour violation (LFV).  An
economical way of radiative neutrino mass generation is to enlarge the
scalar sector of the SM~\cite{Konetschny:1977bn,
  Cheng:1980qt}.  In this work we concentrate on a particularly simple
model of this kind, namely the so-called Zee--Babu
model~\cite{Zee:1985rj,Zee:1985id,Babu:1988ki}. In this model two
$SU(2)_{L}$ singlet scalars are introduced, one singly charged and
one doubly charged, and neutrino masses are generated at two-loop
level.  Through the exchange of heavy scalars,
lepton flavour violating processes such as $\mu \to e\gamma$ can
become observable and the new scalars could be accessible at colliders.
In particular, the doubly charged scalar
may induce very clean like-sign bi-lepton events. Possible connections
to Dark Matter within this model have been discussed in
\cite{Lindner:2011it, Baek:2012ub}.

In this paper we provide an update of previous phenomenological studies
of the Zee--Babu model~\cite{Babu:2002uu, AristizabalSierra:2006gb,
  Nebot:2007bc, Ohlsson:2009vk}, motivated by various new experimental
results relevant for this model. First, precision measurements on
reactor neutrinos~\cite{Abe:2011fz,An:2012eh,Ahn:2012nd} have
confirmed that the smallest neutrino mixing angle is non-vanishing and
close to the previous upper bound, i.e., $\sin^2\theta_{13}\simeq
0.023$~\cite{GonzalezGarcia:2012sz}.  Second, in 2013 the MEG
collaboration has provided a new upper limit on the LFV process
$\mu\to e\gamma$, with a branching ratio less than $5.7\times
10^{-13}$ \cite{Adam:2013mnn}. We perform a parameter scan of the
model taking into account up to date constraints on various LFV and
other low-energy processes as well as neutrino oscillation
experiments. As a consequence we find that most likely the charged
scalars of the Zee--Babu model will be out of reach for the Large
Hadron Collider (LHC), including the 14~TeV configuration. Below we
comment on the possibilities to observe them indirectly through LFV
processes at the proposed International Linear Collider (ILC) in the
like-sign mode.

An alternative way to generate neutrino masses is the so-called
Higgs triplet model, where an $SU(2)_L$ triplet scalar is introduced,
which couples to the lepton doublets and gives rise to a neutrino mass
term from the vacuum expectation value of the neutral
component~\cite{Konetschny:1977bn, Cheng:1980qt,
  Schechter:1980gr,Lazarides:1980nt,Mohapatra:1980yp}.  If the triplet
mass is in the TeV range the doubly charged component could be
produced at colliders through the Drell--Yan process, and subsequently
decay to lepton pairs, leading to similar signatures as the doubly
charged scalar in the Zee--Babu model. If a doubly charged scalar
should be found at a collider below the lower bounds in the Zee--Babu
model discussed below, it may point towards the Higgs triplet
model. In contrast, if no resonance is found the triplet can lead to
similar LFV processes at a like-sign electron collider as the
Zee--Babu scalar. However, due to the different mechanisms to generate
neutrino masses, the specific flavour structure of those processes are
distinctive in the two models. We discuss possibilities to distinguish
the two models, once such LFV events were observed at a future
collider.

The outline of the paper is as follows: In Sec.~\ref{sec:model}, we
present the framework and characteristic features of the Zee--Babu
model. In Sec.~\ref{sec:constraint}, we focus on the low-energy
processes mediated by the doubly charged scalar and summarize the
current constraints on the relevant Yukawa couplings. Numerical
analyses on the model parameters are given in
Sec.~\ref{sec:numerics}. In particular, we illustrate the allowed
ranges of the scalar masses. We further discuss in
Sec.~\ref{sec:colliders} signatures of the doubly charged scalar at a
future linear collider. The discrimination between the Zee--Babu model
and the triplet model is investigated in
Sec.~\ref{sec:type-II}. Finally, in Sec.~\ref{sec:summary}, we
summarize our results and conclude.


\section{The Zee--Babu Model}
\label{sec:model}

The particle content of the Zee--Babu model is that of the SM extended
with two complex $SU(2)_L$ singlet scalars, a singly charged scalar
$h^+$ and a doubly charged scalar $k^{++}$, which couple to
left-handed lepton doublets $L$ and right-handed lepton singlets $e$,
respectively.  The contribution to the Lagrangian is
\begin{eqnarray}\label{eq:L}
{\cal L} & = &  f_{ab} \overline{L^C_{La}} i\sigma_2 L_{Lb} h^{+} + g_{ab} \overline{e^C_a} e_{b}
k^{++} \nonumber \\ && - \mu h^- h^- k^{++} + {\rm h.c.} + V_H \ ,
\end{eqnarray}
where the scalar potential $V_H$ contains additional couplings among scalar fields. The presence of
the tri-linear term $\mu k^{++}h^{-}h^{-}$ together with the two Yukawa-type terms in the first
line of Eq.~\eqref{eq:L} implies that lepton number is violated.\footnote{In the original Zee--Babu
model as displayed in
  Eq.~\eqref{eq:L}, the trilinear term which violates lepton number by
  two units has to be introduced ``by hand''. It is possible to have
  instead a lepton number conserving interaction which generates the
  $\mu$ term by spontaneous symmetry breaking, see e.g.,
  \cite{Chang:1988aa, Lindner:2011it}.}  A Majorana mass term for
neutrinos is generated via a two-loop diagram, yielding
\begin{eqnarray}\label{eq:m}
m^{(\nu)}_{ab} = 16 \mu f_{ac} m_c g^*_{cd} I_{cd} m_d f_{bd} \ ,
\end{eqnarray}
where $m_c$ are charged lepton masses and $I_{cd}$ is a two-loop
integral \cite{McDonald:2003zj}, which approximates to
\begin{equation}\label{eq:I}
I_{cd} \approx I = \frac{1}{(16\pi)^2}\frac{1}{M^2}\frac{\pi^2}{3}
\tilde I \left(\frac{m_k^2}{m_h^2}\right) \,.
\end{equation}
Here, $M = {\rm max}(m_k, m_h)$ and $\tilde I(r)$ is a dimensionless
function of order unity, see e.g., \cite{Nebot:2007bc}.  Note that the
charged scalars couple only to leptons and not at all to
hadrons. Therefore, they might contribute to for instance the Fermi
constant for leptonic processes, and hence lepton--hadron universality
tests provide constraints on the couplings of the scalars, see
section~\ref{sec:constraint}.

Since $f$ is an antisymmetric matrix in flavour space, we have $\det
m^{(\nu)}=0$, and hence one of the light neutrinos is massless. The neutrino mass
eigenvalues $m_1,\,m_2,\,m_3$ are obtained by diagonalization of
\eqref{eq:m} by means of the unitary matrix $U$:
\begin{eqnarray}\label{eq:parametrization}
U  = R_{23}P_{\delta}R_{13}P_{\delta}^{-1}R_{12}P_{M} \ ,
\end{eqnarray}
where $R_{ij}$ correspond to the elementary rotations in the $ij=23$,
$13$, and $12$ planes (parametrized in what follows by three mixing
angles, with $c^{}_{ij} \equiv \cos \theta^{}_{ij}$ and $s^{}_{ij}
\equiv \sin \theta^{}_{ij}$), and $P_{\delta}={\rm diag}(1,1,{\rm
  e}^{{\rm i}\delta})$ and $P_{M}={\rm diag}(1, {\rm e}^{{\rm
    i}\sigma},1)$ contain the Dirac and Majorana CP phases,
respectively.  Here only one Majorana phase $\sigma$ is involved,
since one neutrino is massless.
Depending on the neutrino mass ordering, either $m_1$
(normal ordering, NO) or $m_3$ (inverted ordering, IO) is zero. The
non-zero neutrino mass states are then determined by the solar and
atmospheric mass-squared differences $\Delta m_{21}^2$ and $|\Delta
m_{31}^2|$, where $\Delta m^2_{ij} \equiv m_i^2 - m_j^2$.

Using the antisymmetricity of $f_{ij}$ (the couplings of $h^+$), they
can be expressed in terms of the neutrino mixing angles
\cite{Babu:2002uu, AristizabalSierra:2006gb, Nebot:2007bc}. In the NO case, we have
\begin{eqnarray}\label{eq:eqNH1}
\frac{f_{e\tau}}{f_{\mu\tau}} & = &
\frac{s_{12}c_{23}}{c_{12}c_{13}} + \frac{s_{13}s_{23}}{c_{13}}
e^{-{\rm i}\delta} \ , \\
\label{eq:eqNH2}\frac{f_{e\mu}}{f_{\mu\tau}} & = &
\frac{s_{12}s_{23}}{c_{12}c_{13}} - \frac{s_{13}c_{23}}{c_{13}}
e^{-{\rm i}\delta} \ .
\end{eqnarray}
Since $s_{13}$ is relatively small compared to the other mixing
angles, we can neglect the second terms in the above expressions, and
obtain the approximate relation
\begin{equation}\label{eq:f-NO}
f_{e\mu} \simeq f_{e\tau}\simeq
f_{\mu\tau}/2
\end{equation}
by assuming $s^2_{12} \simeq 1/3$ and $s^2_{23} \simeq
1/2$. For the IO case, the two non-trivial equations are
\begin{eqnarray}\label{eq:eqIH1}
\frac{f_{e\tau}}{f_{\mu\tau}} & = &-\frac{s_{23}c_{13}}{s_{13}}
e^{-{\rm i}\delta} \ , \\
\label{eq:eqIH2} \frac{f_{e\mu}}{f_{\mu\tau}} & = &
\frac{c_{13}c_{23}}{s_{13}} e^{-{\rm i}\delta} \ ,
\end{eqnarray}
which imply
\begin{equation}\label{eq:eqIH3}
\frac{|f_{e\tau}|}{|f_{e\mu}|}  =  \tan\theta_{23} \simeq 1
\quad\text{and}\quad
|f_{\mu\tau}|  \simeq  |f_{e\tau}| \frac{s_{13}}{s_{23}} \ .
\end{equation}

Using Eq.~\eqref{eq:m}, the Yukawa couplings $g_{ab}$ of the doubly charged scalar are related to the neutrino mass matrix
elements as
\begin{flalign}\label{eq:g}
&m^{(\nu)}_{22} = \zeta \left( f^2_{\mu\tau}
\omega_{\tau\tau}-2f_{e\mu}f_{\mu\tau}\omega_{e\tau}+f^2_{e\mu}\omega_{ee}\right)
\nonumber \\
&m^{(\nu)}_{23}  =
\zeta\left(f_{\mu\tau}f_{e\mu}\omega_{e\mu}+f_{e\tau}f_{e\mu}\omega_{ee}-f^2_{\mu\tau}\omega_{\mu\tau}-f_{\mu\tau}f_{e\tau}\omega_{e\tau}\right)
\nonumber \\
&m^{(\nu)}_{33}  =  \zeta \left( f^2_{\mu\tau}
\omega_{\mu\mu}+2f_{e\tau}f_{\mu\tau}\omega_{e\mu}+f^2_{e\tau}\omega_{ee}\right)
\end{flalign}
where $\omega_{ab} = m_a g^*_{ab} m_b$ (no sum) with $m_a$ being the
charged lepton masses and $\zeta \propto \mu$ is a numerical factor
stemming from the loop function.

\section{Experimental Constraints}
\label{sec:constraint}

The experimental bounds on the Zee--Babu model mainly come from
lepton flavour violating processes at low-energy scales mediated by
the heavy scalars, and the universality of weak interactions. In
this section, we summarize the relevant low-energy scale
experimental limits on the Zee--Babu model.

\begin{itemize}

\item Lepton flavour violating decays $\ell^-_a\to\ell^+_b \ell^-_c
\ell^-_d$, which are mediated by the doubly charged scalar $k^{++}$
at tree level. The branching ratio is given by ${\rm
BR}(\ell^-_a\to\ell^+_b \ell^-_c \ell^-_d) = R^{bcd}_a \times {\rm
BR}(\ell^-_a \to \ell^-_b \nu \bar{\nu})$ with
\begin{eqnarray}\label{eq:LFV}
R^{bcd}_a=\frac{1}{2(1+\delta_{cd})}\left|\frac{g_{ab}g^*_{cd}}{G_F
m^2_k}\right|^2 \; .
\end{eqnarray}

\item Universality in $\ell^-_a \to \ell^-_b \nu \bar{\nu}$ decays:
The Fermi coupling constant measured in muon and tau decays obtains
corrections from the exchange of $h^+$, i.e.,
\begin{eqnarray}\label{eq:uni}
\left[\frac{G_{\tau \to \mu}}{G_{\tau \to e}}\right]^2\simeq
1+\frac{\sqrt{2}}{G_F m^2_h} \left( \left|f_{\mu\tau}\right|^2
-\left|f_{e\tau}\right|^2\right) \; .
\end{eqnarray}
Furthermore, by assuming the unitarity of the CKM matrix, one can
test the universality of the couplings in hadronic and leptonic
decays, which gives
\begin{eqnarray}\label{eq:CKM}
|V_{ud}|^2+|V_{us}|^2+|V_{ub}|^2 \simeq 1-\frac{\sqrt{2}}{G_F
m^2_h}\left|f_{e\mu}\right|^2 \; .
\end{eqnarray}
In Eqs.~\eqref{eq:uni} and \eqref{eq:CKM} $G_F$ is the Fermi coupling
constant as given by the SM contribution. We show only
the leading terms in the couplings $f_{ab}$, which emerge from the
interference of the SM diagram with the ones mediated by
the Zee--Babu scalars.

\item Rare lepton decays: $\ell^-_a \to \ell^-_b \gamma$ (for $a\neq
b$) can be mediated at one-loop level by both $k^{++}$ and $h^+$,
and the branching ratios read ${\rm BR}(\ell^-_a\to\ell^+_b \gamma )
= R^{b\gamma}_a \times {\rm BR}(\ell^-_a \to \ell^-_b \nu
\bar{\nu})$, where
\begin{eqnarray}\label{eq:rare}
R^{b\gamma}_a= \frac{\alpha}{48\pi}\left( \left|\frac{(f^\dagger f)_{ab}}{G_F m^2_h}\right|^2 +
16\left|\frac{(g^\dagger g)_{ab}}{G_F m^2_k}\right|^2\right) \; .
\end{eqnarray}

\item Muonium to antimuonium conversion through the exchange of $k^{++}$:
The process $\mu^+ e^- \to \mu^- e^+$ is well bounded experimentally, leading to constraints on
the effective coupling related to the following four-fermion
operator
\begin{eqnarray}\label{eq:muonium}
G_{M\bar{M}} = -\frac{\sqrt{2}}{8}\frac{g_{ee}g^*_{\mu\mu}}{m_k^2}
\, .
\end{eqnarray}

\item Muon and electron anomalous magnetic moments: $a=(g-2)/2$
obtains addition contributions $\delta a$ from both
$h^+$ and $k^{++}$, with
\begin{eqnarray}
\delta a_a =-\frac{m_a^2}{24\pi^2}\left(\frac{(f^\dagger
f)_{aa}}{m_h^2}+4\frac{(g^\dagger g)_{aa}}{m_k^2}\right)\,,
\end{eqnarray}
where $a=e,\mu$. The bound from $\delta a_e$ is very weak (only
relevant for scalar masses above $10^3$~TeV) and therefore we include
only the constraint from $\delta a_\mu$.

\item $\mu - e$ conversion in nuclei: The loops which mediate the decays $\mu^-\rightarrow e^-\gamma$ generate an effective $\mu e\gamma$ vertex which  induces $\mu-e$ conversion in
nuclei. Using the result from \cite{Kitano:2002mt} we obtain
\begin{align}
{\rm CR}&(\mu N \to e N) \simeq \frac{2e^2G^2_F}{\Gamma_{\rm
capt}}\times \\
&\left(\left|A^h_R D +e A^h_L V^{(p)}\right|^2
+\left|A^k_R D +e A^k_L V^{(p)}\right|^2 \right) \,, \nonumber
\end{align}
where $D$ and $V^{(p)}$ represent overlap integrals of the muon and
electron wave functions. The form factors are given by the same expressions as
in the case of the Higgs triplet model \cite{Dinh:2012bp}
\begin{eqnarray}
A^h_R &=& -\frac{\left(f^\dagger f\right)_{e\mu}}{768\sqrt{2}\pi^2 G_F m^2_h} \; , \nonumber \\
A^k_R &=& -\frac{\left(g^\dagger g\right)_{e\mu}}{48\sqrt{2}\pi^2 G_F m^2_k} \; , \nonumber \\
A^h_L &=& -\frac{\left(f^\dagger f\right)_{e\mu}}{144\sqrt{2}\pi^2 G_F m^2_h} \; , \nonumber \\
A^k_L &=& -\sum_{a=e,\mu,\tau}
\frac{g^*_{a e} g_{a \mu}}{6\sqrt{2}\pi^2 G_F m^2_k} F\left(\frac{-q^2}{m^2_k},\frac{m^2_a}{m^2_k}\right) \; ,
\label{eq:m2e-conv}
\end{eqnarray}
where the loop function is~\cite{Ma:2000xh}
\begin{eqnarray}
F(x,y)& =& \frac{4y}{x}+\log(y) + \left(1-\frac{2y}{x}\right)
\nonumber
\\ &\times & \sqrt{1+\frac{4y}{x}}\log\frac{\sqrt{x+4y}+\sqrt{x}}{\sqrt{x+4y}-\sqrt{x}}
\; .
\end{eqnarray}
Note that in the Higgs triplet model both the singly and doubly
charged scalars couple to left-handed leptons (since both are
components of the same $SU(2)$ triplet field), whereas in the Zee--Babu
model $h$ couples to left-handed and $k^{++}$ couples to right-handed
leptons, see Eq.~\eqref{eq:L}. Therefore, the amplitudes for singly
and doubly charged scalar mediated processes do not interfere in the
case of the Zee--Babu model, whereas they do in the case of the Higgs
triplet model~\cite{Kitano:2002mt}.
\end{itemize}

\begin{table*}[t]
\begin{tabular}{lcc|r} \hline\hline Constraint & &Ref.&
Bound ($90\%$ C.L.)\\ \hline
$\sum_{q=d,s,b}|V_{uq}|^2$ &$0.99990\pm0.0006$&\cite{Beringer:1900zz}&$|f_{e\mu}|^2<0.014\left(\frac{m_h}{\rm{TeV}}\right)^2$\\
$\mu-e$ universality & $\frac{G_{\tau \to \mu}}{G_{\tau \to
e}}=1.0001\pm0.0020$ & \cite{Beringer:1900zz} & $\left|
\left|f_{\mu\tau}\right|^2
-\left|f_{e\tau}\right|^2\right|<0.05\left(\frac{m_h}{\rm{TeV}} \right)^2$\\
$\mu-\tau$ universality & $\frac{G_{\tau \to e}}{G_{\mu \to
e}}=1.0004\pm0.0022$ & \cite{Beringer:1900zz} & $\left|
\left|f_{e\tau}\right|^2
-\left|f_{e\mu}\right|^2\right|<0.06\left(\frac{m_h}{\rm{TeV}} \right)^2$\\
$e-\tau$ universality & $\frac{G_{\tau \to \mu}}{G_{\mu \to
e}}=1.0004\pm0.0023$ & \cite{Beringer:1900zz} & $\left|
\left|f_{\mu\tau}\right|^2
-\left|f_{e\mu}\right|^2\right|<0.06\left(\frac{m_h}{\rm{TeV}} \right)^2$\\
$\delta a_\mu$ &$(28.7 \pm 80 )\times10^{-10}$&\cite{Bennett:2006fi,Beringer:1900zz}&  $ r(|f_{e\mu}|^2+|f_{\mu\tau}|^2) + 4 (|g_{e\mu}|^2+|g_{\mu\mu}|^2+|g_{\mu\tau}|^2)<3.4 \left(\frac{m_k}{\rm{TeV}} \right)^2$ \\
$\mu^-\rightarrow e^+e^-e^-$& BR$<1.0\times10^{-12}$&\cite{Bellgardt:1987du}&$|g_{e\mu}g_{ee}^\ast|<2.3\times10^{-5}\left(\frac{m_k}{\rm{TeV}}\right)^2$\\
$\tau^-\rightarrow e^+e^-e^-$ &BR$<2.7\times10^{-8}$&\cite{Hayasaka:2010np}&$|g_{e\tau}g_{ee}^\ast|<0.009\left(\frac{m_k}{\rm{TeV}}\right)^2$\\
$\tau^-\rightarrow e^+e^-\mu^-$ &BR$<1.8\times10^{-8}$&\cite{Hayasaka:2010np}&$|g_{e\tau}g_{e\mu}^\ast|<0.005\left(\frac{m_k}{\rm{TeV}}\right)^2$\\
$\tau^-\rightarrow e^+\mu^-\mu^-$ &BR$<1.7\times10^{-8}$&\cite{Hayasaka:2010np}&$|g_{e\tau}g_{\mu\mu}^\ast|<0.007\left(\frac{m_k}{\rm{TeV}}\right)^2$\\
$\tau^-\rightarrow \mu^+e^-e^-$ &BR$<1.5\times10^{-8}$&\cite{Hayasaka:2010np}&$|g_{\mu\tau}g_{ee}^\ast|<0.007\left(\frac{m_k}{\rm{TeV}}\right)^2$\\
$\tau^-\rightarrow \mu^+e^-\mu^-$ &BR$<2.7\times10^{-8}$&\cite{Hayasaka:2010np}&$|g_{\mu\tau}g_{e\mu}^\ast|<0.006\left(\frac{m_k}{\rm{TeV}}\right)^2$\\
$\tau^-\rightarrow \mu^+\mu^-\mu^-$ &BR$<2.1\times10^{-8}$&\cite{Hayasaka:2010np}&$|g_{\mu\tau}g_{\mu\mu}^\ast|<0.008\left(\frac{m_k}{\rm{TeV}}\right)^2$\\
$\mu\rightarrow e\gamma$ &BR$<5.7\times10^{-13}$&\cite{Adam:2013mnn}& $r^2 |f^*_{e\tau} f_{\mu\tau}|^2+16| g^*_{e\alpha}g_{\alpha \mu}|^2<1.6\times10^{-6}\left(\frac{m_k}{\rm{TeV}}\right)^4$\\
$\tau\rightarrow e\gamma$ &BR$<3.3\times10^{-8}$&\cite{Aubert:2009ag}&$r^2 |f^*_{e\mu} f_{\mu\tau}|^2+16| g^*_{e\alpha}g_{\alpha \tau}|^2<0.52\left(\frac{m_k}{\rm{TeV}}\right)^4$ \\
$\tau\rightarrow\mu\gamma$&BR$<4.5\times10^{-8}$ &\cite{Aubert:2009ag}&$r^2 |f^*_{e\mu} f_{e\tau}|^2+16| g^*_{\mu\alpha}g_{\alpha \tau}|^2<0.71\left(\frac{m_k}{\rm{TeV}}\right)^4$ \\
$\mu \leftrightarrow e$ conversion &${\rm CR}<7.0\times 10^{-13}$ & \cite{Bertl:2006up}
&
see Eq.~\eqref{eq:m2e-conv}
\\
$\mu^+ e^- \to \mu^- e^+$ &  $G_{M\bar{M}}<3\times 10^{-3} G_F$ & \cite{Beringer:1900zz} & $|g_{ee}g^*_{\mu\mu}|<0.2\left(\frac{m_k}{\rm{TeV}}\right)^2$ \\
\hline\hline
\end{tabular}
\caption{\label{tab:1} Summary of experimental
constraints and the corresponding bounds on the Yukawa couplings.
Here $r=m^2_k/m^2_h$, and  $g^*_{e\alpha} g_{\alpha \mu} = g^*_{e e}
g_{e \mu} + g^*_{e \mu} g_{\mu \mu} +g^*_{e \tau} g_{\tau \mu}$ and
so on.}
\end{table*}

We summarize in Table~\ref{tab:1} the low-energy experimental
constraints used in our analysis.  One can observe that lepton flavour
violating processes set more stringent bounds on the Yukawa couplings,
in particular the $\mu\to e \gamma$ and $\mu \to 3 e$ decays. The
later process could however be suppressed in the Zee--Babu model if
$g_{ee}$ or $g_{e\mu}$ is vanishing, which is possible while still
obtaining a valid neutrino mass matrix.  The $\mu \to e \gamma $ decay
is mediated by both singly and doubly charged scalars, and is
proportional to both Yukawa couplings $f$ and $g$, which cannot
vanish simultaneously.  Therefore, the most stringent constraint on
the Zee--Babu model stems from the $\mu \to e \gamma$ decay.

\section{Numerical Analysis}
\label{sec:numerics}

We perform a scan of the Zee--Babu model parameters confronting the
experimental data in order to obtain constraints on the scalar
masses. The independent parameters can be chosen as: three leptonic
mixing angles; Dirac and Majorana phases $\delta$, $\sigma$; the
Yukawa couplings $g_{ee}$, $g_{e\mu}$, $g_{e\tau}$, $f_{\mu\tau}$;
scalar masses $m_k$, $m_h$; and the $\mu$ parameter in the scalar
potential. Neutrino masses are fixed to the values set by the best fit
mass-squared differences~\cite{GonzalezGarcia:2012sz}. Then the
remaining Yukawa couplings $g_{ab}$ and $f_{ab}$ are fixed by
Eqs.~\eqref{eq:eqNH1}, \eqref{eq:eqNH2}, \eqref{eq:eqIH1},
\eqref{eq:eqIH2}, \eqref{eq:g}. For simplicity we fix the mixing
angles to the values \cite{GonzalezGarcia:2012sz} $\sin^2\theta_{12} =
0.30$, $\sin^2\theta_{23} = 0.41$, $\sin^2\theta_{13} = 0.023$. The
remaining parameters are scanned in the following ranges:
\begin{eqnarray}\label{eq:p}
\delta & = & [0,2\pi) \nonumber \\
\sigma & = & [0,\pi) \nonumber \\
|g_{ee}|, \ |g_{e\mu}|, \ |g_{e\tau}|, \  f_{\mu\tau} & =& [0,\kappa) \nonumber \\
\mu & =& [0,\lambda \times \min(m_k,m_h))
\end{eqnarray}
where $\kappa$ parametrizes the requirement of perturbativity of
Yukawa couplings. If not stated otherwise we take
$\kappa=1$. For the tri-linear term, the $\mu$ parameter
  induces loop corrections to the scalar masses as $\delta m^2_{k,h}
  \sim \mu^2/(4\pi)^2$. In the absence of fine-tuning the correction
  should be smaller than the tree-level masses, which leads to the
  constraint $\mu \ll 4\pi m_{k,h}$. Henceforth, we parameterize this
  requirement by a parameter $\lambda$, see Eq.~\eqref{eq:p}. The
phases of $g_{ee}, \ g_{e\mu}, \ g_{e\tau}$ are chosen randomly,
whereas $f_{\mu\tau}$ can be taken real without loss of
generality. For a given set of the parameters in Eq.~\eqref{eq:p} we
check if all other values for $g_{ab}$ and $f_{ab}$ are less than
$\kappa$; if not then the point is discarded. If the perturbativity
constraint is fulfilled we compare the model predictions to the
experimental data with a $\chi^2$ function
\begin{eqnarray}\label{eq:chi2}
\chi^2_i = \frac{(\rho_i - \rho^0_i)^2}{\sigma^2_{i}} \ ,
\end{eqnarray}
where $\rho^0_i$ represents the data of the $i$th experimental
observable, $\sigma_i$ the corresponding 1$\sigma$ absolute error, and
$\rho_i$ the prediction of the model. The index $i = 1,...,17$ runs over the 17
experimental observables given in Table~\ref{tab:1}. In case of upper
bounds we set $\rho^0_i = 0$ and use the $1\sigma$ upper bound for
$\sigma_i$. In order to identify the allowed regions in parameter space we proceed as follows.
For a given point in parameter space we consider the maximum $\chi^2_i$ of all data points:
\begin{equation}
  \chi^2_{\rm max} = \max_i \chi^2_i \,.
\end{equation}
If $\chi^2_{\rm max} \le 4$ is fulfilled we keep the point, otherwise it is
discarded. In that way we make sure that all data points are fitted
within 2 standard deviations. Let us stress that we do not adopt any
particular statistical interpretation of the resulting regions in
parameter space in terms of confidence regions, apart from the above
statement that all constraints are satisfied within $2\sigma$.

From Eqs.~\eqref{eq:g} one can see that the contribution of the
couplings $g_{ee}$ ($g_{e\mu}$, $g_{e\tau}$) is suppressed by two
powers (one power) of the electron mass. Indeed, we find always viable
solutions for $g_{ee} = g_{e\mu} = g_{e\tau} = 0$. However, we do take
into account finite values in our scan in order to allow for
sub-leading effects induced by those couplings.

\begin{figure*}[!t]
\begin{center}  
\includegraphics[width=0.45\textwidth]{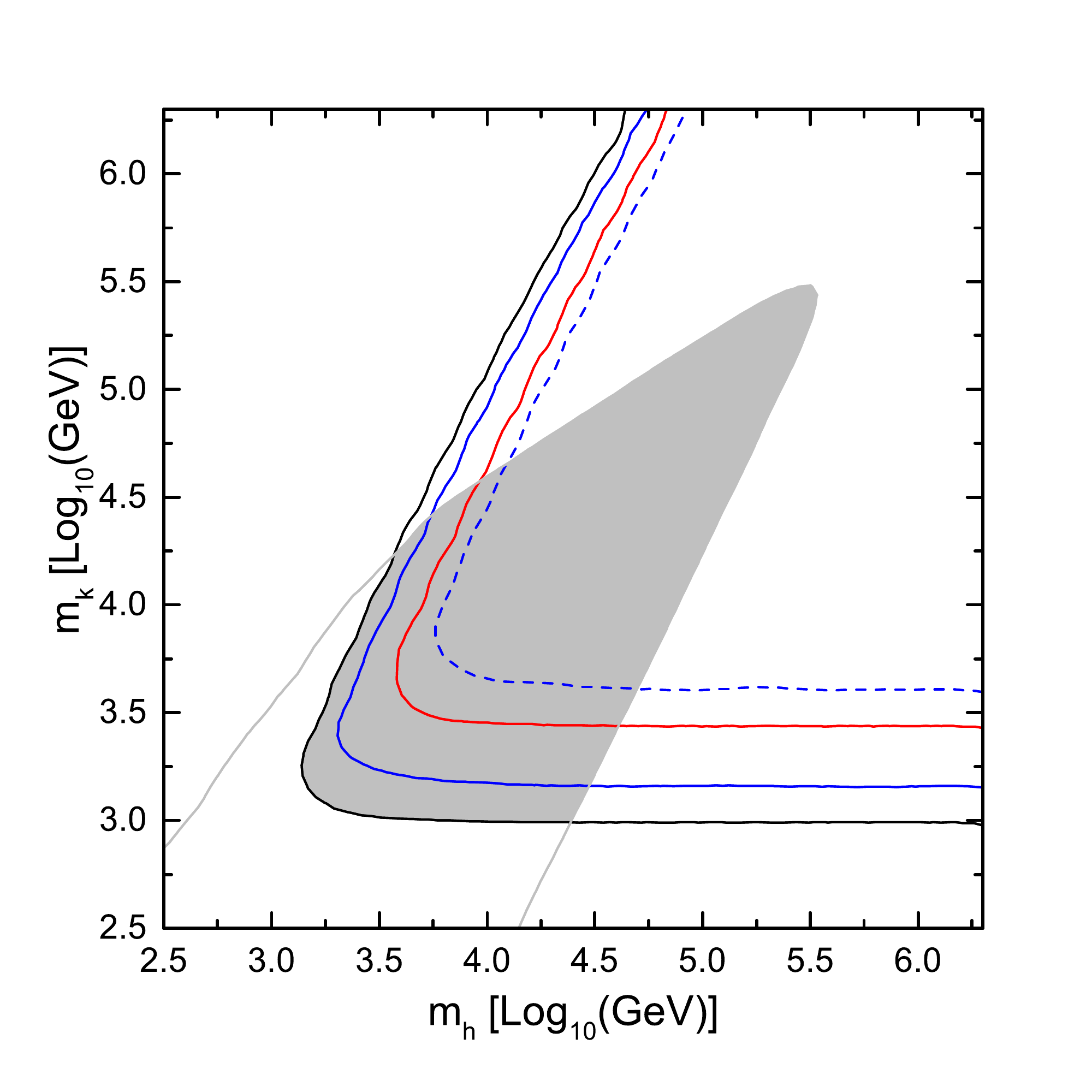}
\includegraphics[width=0.45\textwidth]{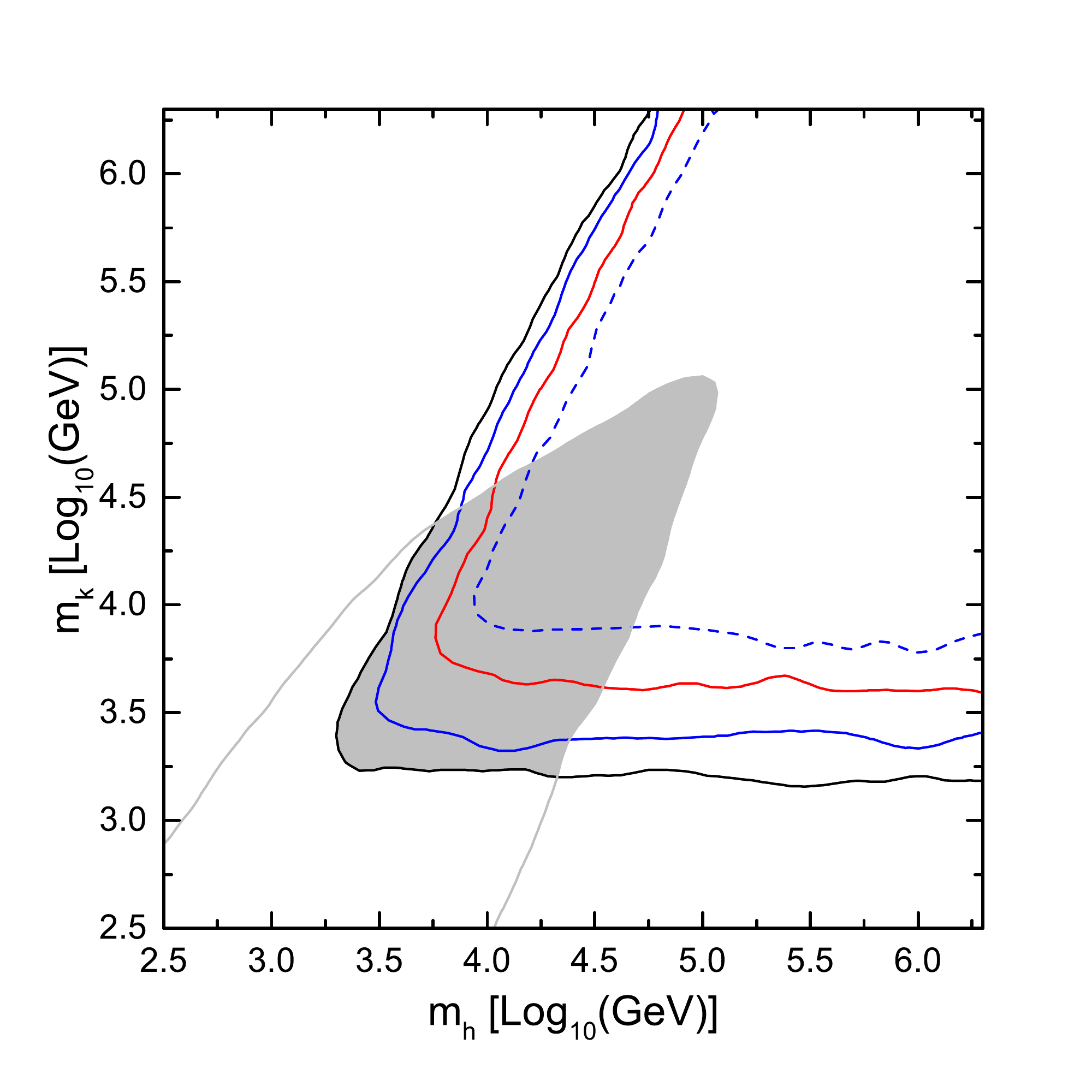}
\caption{\label{fig:fig1} The shadowed regions correspond to allowed
  ranges of the scalar masses for the normal mass ordering (left
  panel) and the inverted mass ordering (right panel) by requiring
  $\chi^2_{\rm max}<4$ and imposing the perturbativity criterion
  $\lambda=\kappa = 1$. The black, blue, and red curves correspond to lower
  limits on the scalar masses obtained from the experimental data by
  requiring $\chi^2_{\rm max}<4$ but without the $\kappa$
  constraint. The black curve corresponds to the current experimental
  bounds. The blue solid and blue dashed curves show the exclusion regions from
    the expected $\mu-e$ conversion constraint ${\rm CR}(\mu {\rm Al}
    \to e {\rm Al})<6\times 10^{-17}$ and ${\rm CR}(\mu {\rm Al}
    \to e {\rm Al})<10^{-18}$, respectively. The red line is given
    by assuming BR$(\mu\rightarrow e\gamma)<10^{-14}$.  Furthermore, gray curves delimit
  the region allowed by perturbativity without requiring that the
  experimental constraints are respected.}
\end{center}
\end{figure*}

The allowed ranges of scalars masses are illustrated in Fig.~\ref{fig:fig1} for both normal and
inverted mass orderings.  For scalar masses within the shadowed regions all the constraints are
satisfied in the sense of $\chi^2_{\rm max}< 4$ and $\lambda=\kappa=1$ as explained above.  We
observe that the parameter space of the model is closed, however, allowing for $\lambda\sim \kappa
\sim 1$ scalar masses up to $\mathcal{O}$(100~TeV) are possible. If we allow for some fine-tuning
in the scalar potential by setting the parameter $\lambda$ to $4\pi$, the upper bound on the scalar
masses will be larger than $10^3$~TeV (see Fig.~\ref{fig:lambda5} below).  Note however, that large
scalar masses require Yukawa couplings close to the perturbativity limit. If Yukawa couplings
assume values $g_{\alpha\beta}, f_{\alpha\beta} \ll 1$ the scalar masses get pushed towards lower
values. The lower bound on the scalar masses (black curve) is dominated by the observables from
Tab.~\ref{tab:1}, most importantly from the MEG bound on $\mu\to e\gamma$. We obtain the following
lower bounds by requiring that all constraints are satisfied at $2\sigma$:
\begin{align}\label{eq:massbound}
\begin{split}
  &m_k > 1.3~{\rm TeV} \,,\quad
  m_h > 1.3~{\rm TeV} \quad\text{(NO)} \\
  &m_k > 1.9~{\rm TeV} \,,\quad
  m_h > 2.0~{\rm TeV} \quad\text{(IO)} \\
\end{split}
\quad (\lambda = 1).
\end{align}
Note that the lowest possible value for the doubly charged scalar
$k$ occurs for relatively large values of singly charged scalar mass
and depends also on the perturbativity/fine-tuning conditions. In
deriving the bounds we have assumed $\lambda = 1$. If we allow values
for the tri-linear coupling $\mu$ larger than the scalar masses
(amounting to some fine-tuning in the scalar potential) the lower
bounds on the Zee--Babu scalars can be relaxed. For example, we show
the mass ranges in Fig.~\ref{fig:lambda5} by taking a relatively large
constraint $\lambda=5$. One can see that the lower bounds on the
scalar masses reduce to
\begin{align}\label{eq:massbound-5}
\begin{split}
  &m_k > 0.5~{\rm TeV} \,,\quad
  m_h > 0.6~{\rm TeV} \quad\text{(NO)} \\
  &m_k > 0.8~{\rm TeV} \,,\quad
  m_h > 1.0~{\rm TeV} \quad\text{(IO)} \\
\end{split}
\quad(\lambda=5).
\end{align}
The bounds for IO can be further weakened by fine tuning of the
complex phases $\delta \approx \pi$ and $\sigma \approx \pi/2$,
leading to a cancellation between different terms in Eq.~\eqref{eq:g}.
By performing a dedicated search with phases constrained to be very
close to those special values we find that the IO bounds for $m_k$ and
$m_h$ in Eq.~\eqref{eq:massbound} reduce to around 1.0 and 1.1~TeV,
respectively. Let us also stress that our bounds are obtained by a
random parameter scan, throwing $10^{5}$ points for a given choice of
$m_k$ and $m_h$, out of which only a fraction passes our
perturbativity requirement. With such a method fine tuned solutions as
the one mentioned above might be missed. In this sense our bounds for
the scalar masses hold for ``generic'' values of the parameters.

\begin{figure*}[t]
\begin{center}  
\includegraphics[width=0.45\textwidth]{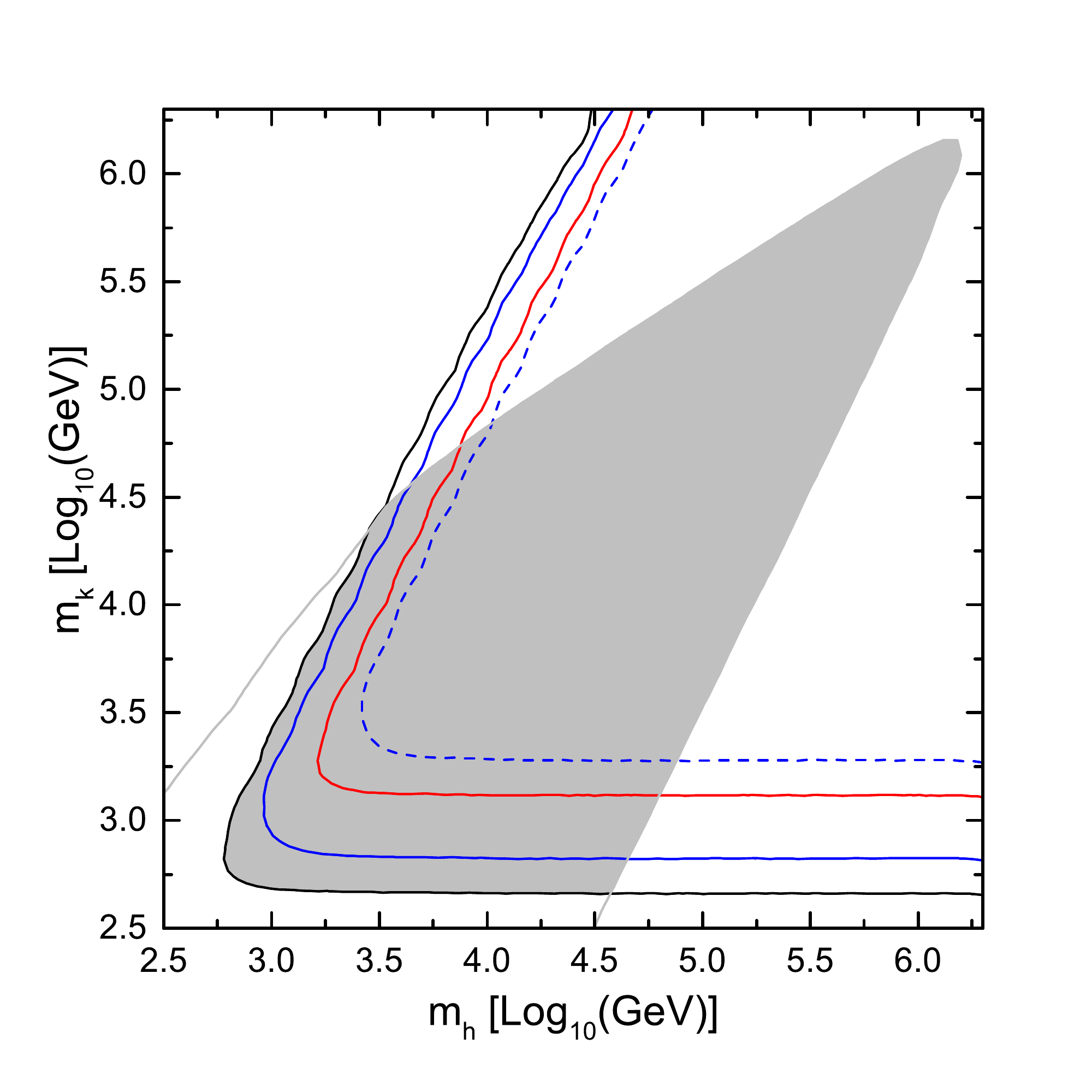}
\includegraphics[width=0.45\textwidth]{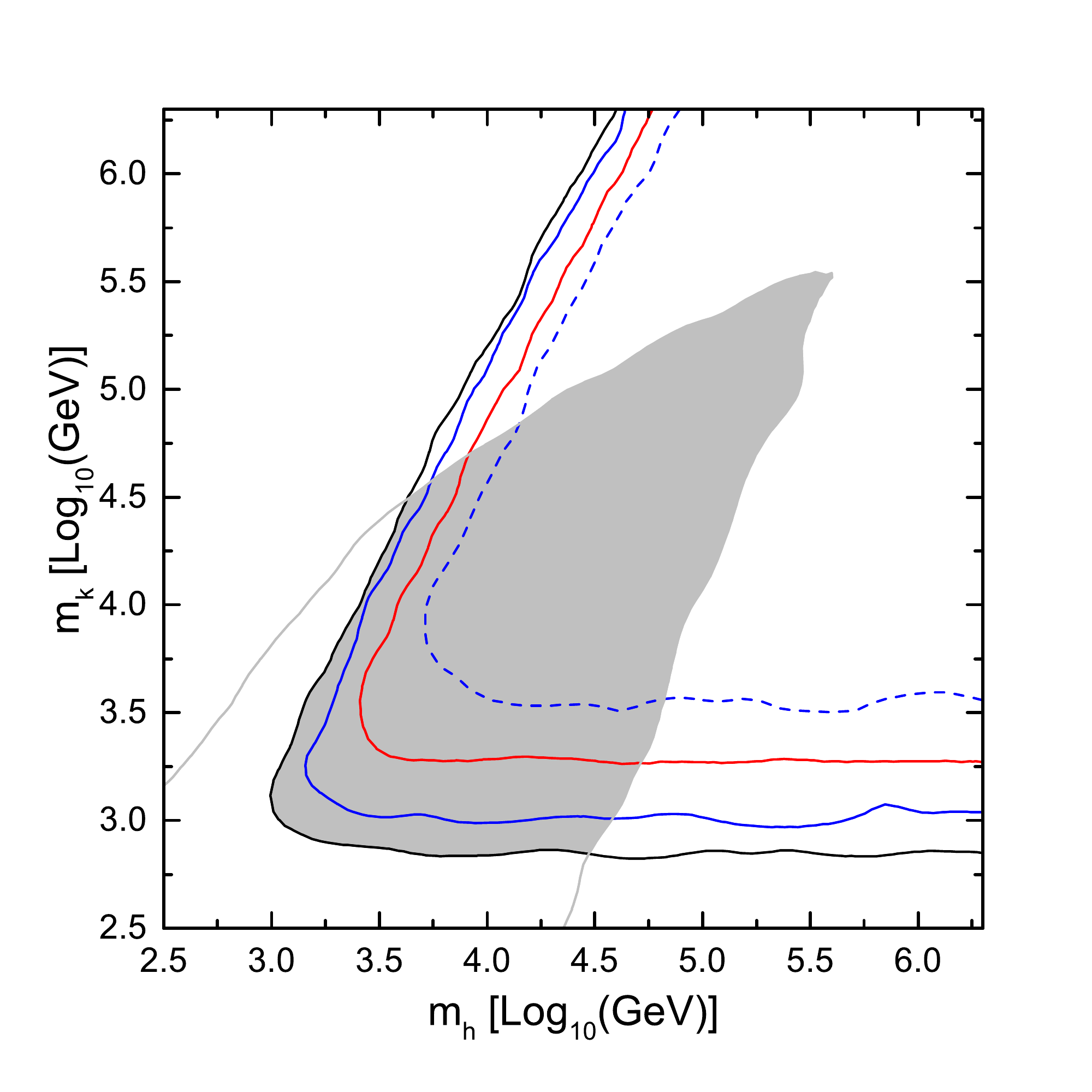}
\caption{\label{fig:lambda5} The same as Fig.~\ref{fig:fig1} with $\lambda=5$, i.e., allowing for a tri-linear coupling $\mu$ larger than the scalar masses, see Eq.~\eqref{eq:p}.}
\end{center}
\end{figure*}

There are several future projects aiming for improving significantly
the bound on $\mu - e$ conversion by about 4 to 5 orders of magnitude
compared to the current limit \cite{Bertl:2006up}, for instance the
Mu2e \cite{Abrams:2012er, Knoepfel:2013nqa} and COMET
\cite{Cui:2009zz} experiments aim at sensitivities of order
$10^{-16}$, whereas the target sensitivity of the PRISM
project~\cite{Witte:2012zza} is even of order $10^{-18}$. Furthermore,
an upgrade program is underway for the MEG experiment aiming at a
sensitivity improvement of a further order of
magnitude~\cite{Adam:2013mnn}. We thus also show in
Figs.~\ref{fig:fig1} and \ref{fig:lambda5} with colored contours the
future experimental constraints on the charged scalars. Those improved
constraints on LFV (if no positive signal is found) will further push
up the lower bounds on the scalar masses of about 1 order of
magnitude. However, it is still difficult to entirely rule out the
Zee--Babu model, no matter for the normal or inverted mass ordering
(depending on the perturbativity requirements).

\begin{figure}[!t]
\begin{center}  
\includegraphics[width=0.45\textwidth]{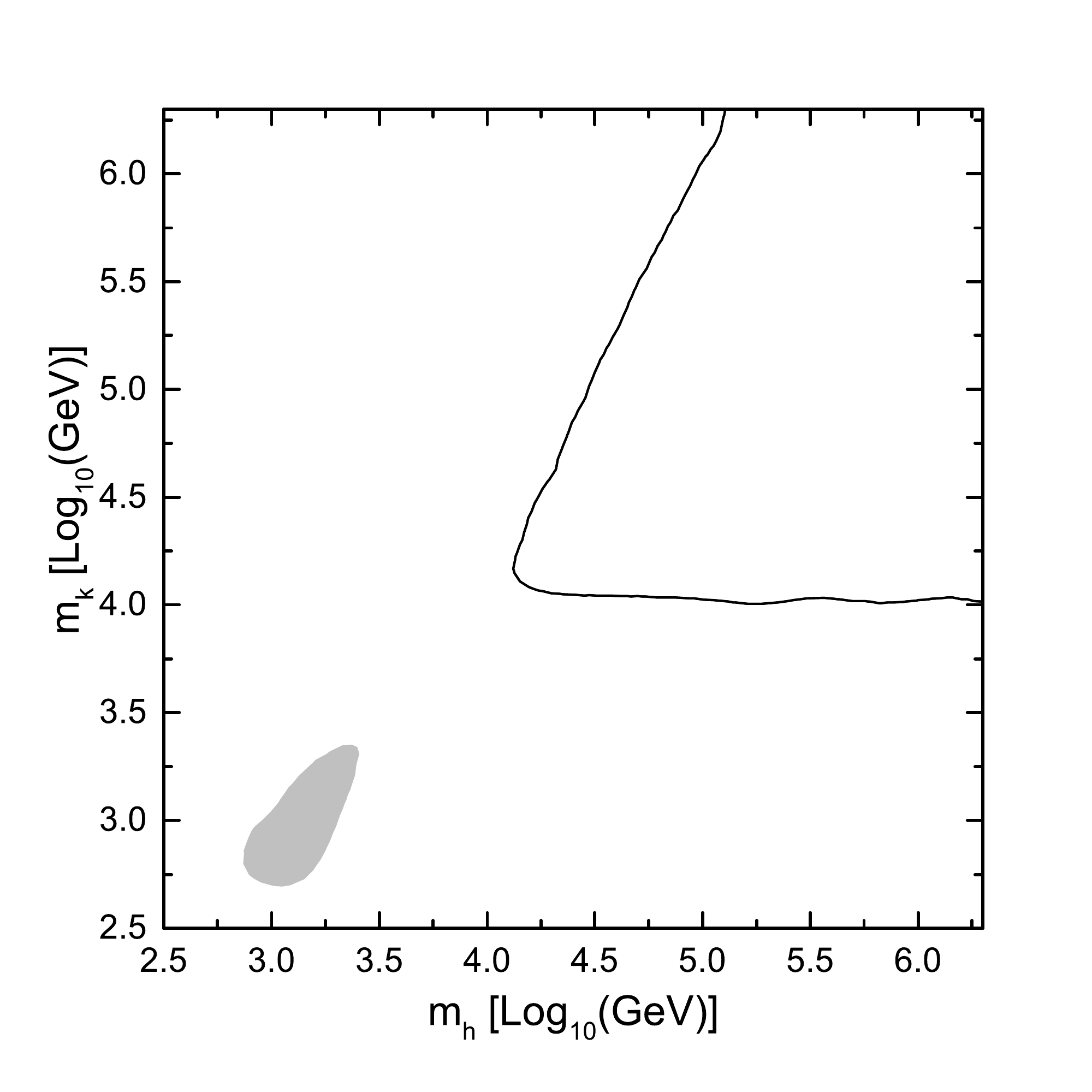} 
\caption{\label{fig:th13small} Allowed ranges of the scalar masses for
  the inverted mass ordering for $s_{13} = 0.001$ by requiring
  $\chi^2_{\rm max} < 4$. The black curve corresponds to the current
  experimental bounds. The shadowed area contour is allowed by the
  perturbativity criterion $\lambda=\kappa = 1$ (without imposing the
  constraints from Tab.~\ref{tab:1}).}
\end{center}
\end{figure}

It should be noticed that our analysis is based on the latest
measurement on the smallest mixing angle $\theta_{13}$, which is
indeed very crucial for the IO case. For illustration purposes, we
show in Fig.~\ref{fig:th13small} also a similar numerical analysis but
using $s^2_{13} =0.001$ (which is by now excluded by oscillation
data). In this case, there is no overlap between the perturbativity
contours and the experimental contours, indicating that the Zee--Babu
model is incompatible with the inverted mass ordering if $\theta_{13}$
is very small. In other words, if the neutrino mass ordering is
inverted, the Zee--Babu model can then be viewed as a natural
candidate for predicting sizeable $\theta_{13}$, see
e.g.,~\cite{Nebot:2007bc}.  Such a feature is also manifest in
Eqs.~\eqref{eq:eqIH1} and \eqref{eq:eqIH2}, showing that both, $f_{e\mu}$ and
$f_{e\tau}$ are inversely proportional to $s_{13}$, and therefore a
too small $s_{13}$ may blow up Yukawa couplings resulting in a
conflict with lepton flavour violation constraints. Conversely, there
is no inverse dependence on $\theta_{13}$ in the case of normal mass
ordering, c.f.\ Eqs.~\eqref{eq:eqNH1} and \eqref{eq:eqNH2}, and thus
the choice of $\theta_{13}$ has no important impact on the Zee--Babu
model parameters. We have checked this point numerically by choosing
different values for $\theta_{13}$, and they all give almost the same
scalar mass ranges for normal ordering.

\bigskip

As discussed above, present data pushes the masses of the singly and
doubly charged scalars of the Zee--Babu model above the TeV, see
Eq.~\eqref{eq:massbound}. This makes the direct production at
colliders difficult. The production cross sections of the Zee--Babu
scalars have been calculated in~\cite{Babu:2002uu,
  AristizabalSierra:2006gb, Nebot:2007bc}, for recent work on
di-leptons in general see e.g., \cite{delAguila:2013yaa,
  Alloul:2013raa}. ATLAS~\cite{ATLAS:2012hi} and
CMS~\cite{Chatrchyan:2012ya} have searched for doubly charged scalars decaying predominantly into
muons and/or electrons based on approximately 5~fb$^{-1}$ at $\sqrt{s} = 7$~TeV, obtaining lower
bounds on their mass of around 400~GeV. The results of \cite{Nebot:2007bc} show that for the
Zee--Babu doubly charged scalar with masses $m_k = 0.5, \, 1, \, 1.5$~TeV of order 300, 10, 1
events are expected at LHC for 300~fb$^{-1}$ at $\sqrt{s} = 14$~TeV, respectively, assuming 100\%
branching fraction of $k^{++}$ into leptons. Given the constraints on the masses derived above this
implies that most likely the Zee--Babu scalars will not be observable at LHC, unless some degree of
fine-tuning is accepted and the relaxed bounds of Eq.~\eqref{eq:massbound-5} apply. In that case,
future experiments for LFV would also observe a positive signal, see Fig.~\ref{fig:lambda5}.

\section{Tests at an $e^- e^-$ collider}
\label{sec:colliders}

At a possible future $e^+ e^-$ linear collider the scalars can be
pair-produced by photon and $Z$ exchange: $e^+ e^- \to k^{++}
k^{--}$. Obviously this requires center of mass energies of $\sqrt{s}
> 2 m_{k,h}$, which in view of the bounds in Eq.~\eqref{eq:massbound}
seems not realistic in the foreseeable future. However, a linear
collider may also be operated in the like-sign mode. This offers a new
window to search for LFV processes within the context of the Zee--Babu
model for scalar masses up to $\sim 10$~TeV.

The possibility to test bi-leptons at a like-sign electron collider
has been considered since long time, see Refs.~\cite{Rizzo:1981xx,
  Rizzo:1982kn, London:1987nz, Cuypers:1997qg, Raidal:1997tb,
  Cakir:2006pa, Rodejohann:2010jh, Rodejohann:2010bv} for an
incomplete list of references. Lots of work has been devoted to the search for lepton number
violating reactions $e^- e^- \to W^- W^-$ in the context of Higgs triplet models. In the Zee--Babu
model this process is not allowed at tree level. However, lepton flavour violating (but lepton
number conserving) reactions such as $e^-e^- \to \ell^-_{\alpha} \ell^-_{\beta}$ mediated by
$k^{--}$ at tree level may be observable.

\begin{figure}[t]
\begin{center}
\includegraphics[width=8.5cm]{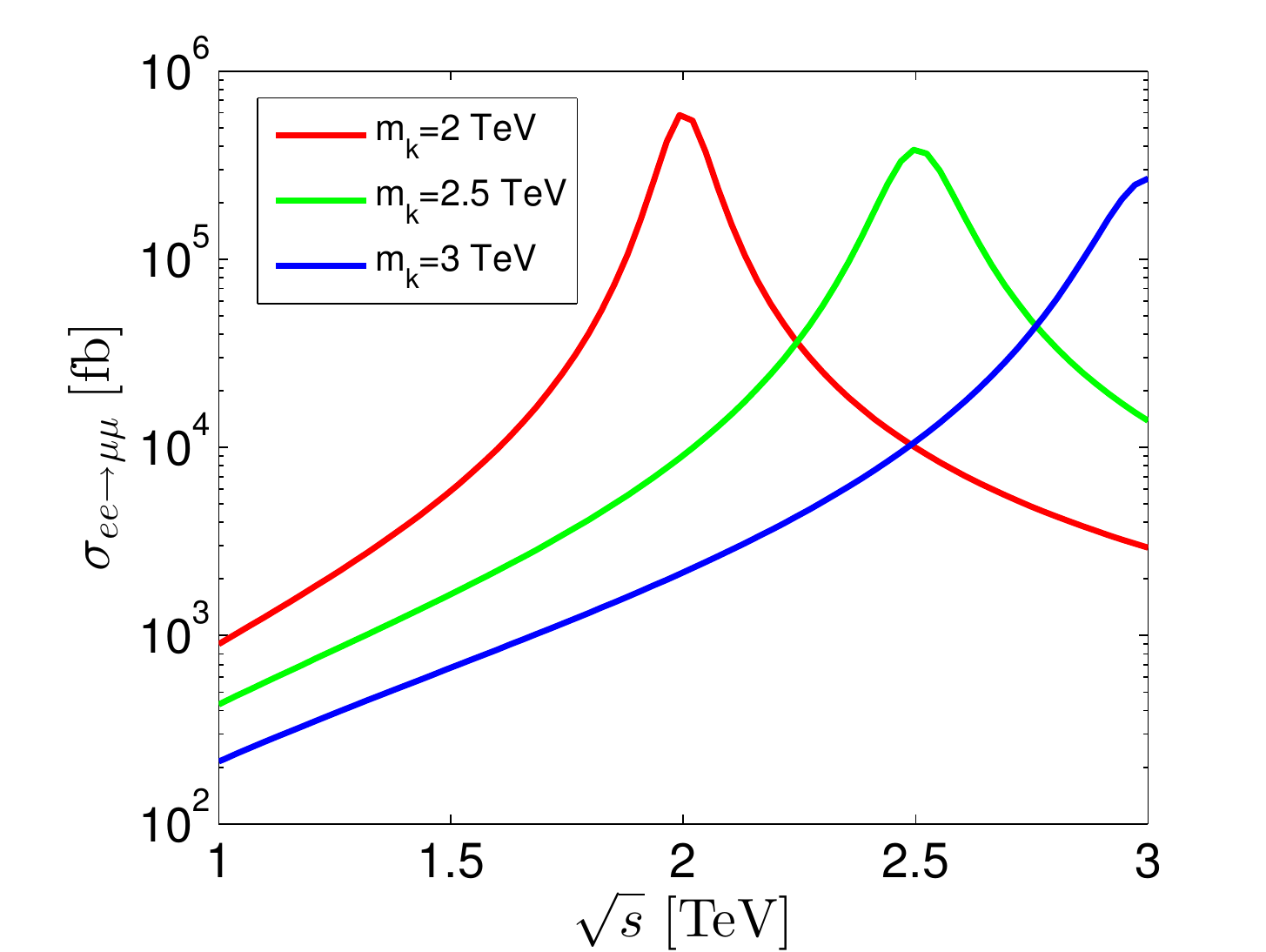}
\caption{\label{fig:figEE} Upper limit on the cross section
  $\sigma(e^-e^-\to \mu^-\mu^-)$ as a function of the center of mass
  energy $\sqrt{s}$ for normal neutrino mass ordering for doubly
  charged scalar masses of 2, 2.5, and 3~TeV.}
\end{center}
\end{figure}

The cross section for the process $e^- e^- \to \ell^-_\alpha \ell^-_\beta$ for
$(\alpha\beta) \neq (ee)$ is given by
\begin{align}\label{eq:XS-likesign}
  \sigma(ee \to \alpha\beta) = \frac{S|g_{ee} g_{\alpha\beta}|^2}{4\pi(1+\delta_{\alpha\beta})}
  \frac{s}{(s - m_k^2)^2 + m^2_k \Gamma_k^2} \,,
\end{align}
where $\Gamma_k$ is the width of the doubly charged scalar, and
$S=(1+P_1)(1+P_2)$ the polarization factor of the incoming electron
beams. First, we note that this process is proportional to
$|g_{ee}|^2$. As mentioned above, this coupling is not determined by
neutrino data and in principle it could be zero.  Hence, no signal can
be predicted. On the other hand, also the upper bound on this coupling
is rather weak and therefore a sizeable cross section would be
possible in principle. The upper limit on the cross section as a
function of the center of mass energy is shown in Fig.~\ref{fig:figEE}
for unpolarized beams ($S=1$).
For center of mass energies of $\sqrt{s} > m_k$ a sharp resonance can
be observed, leading to very large cross sections in excess of 100~pb,
allowing for the direct discovery of the doubly charged scalar. In
view of the bounds from Eq.~\eqref{eq:massbound}, this will require a
multi-TeV collider. However, for lower center of mass energies, one
may still expect visible cross sections, corresponding to contact
interactions mediated by the heavy scalar.

\begin{figure*}[!t]
\begin{center}
\includegraphics[width=0.45\textwidth]{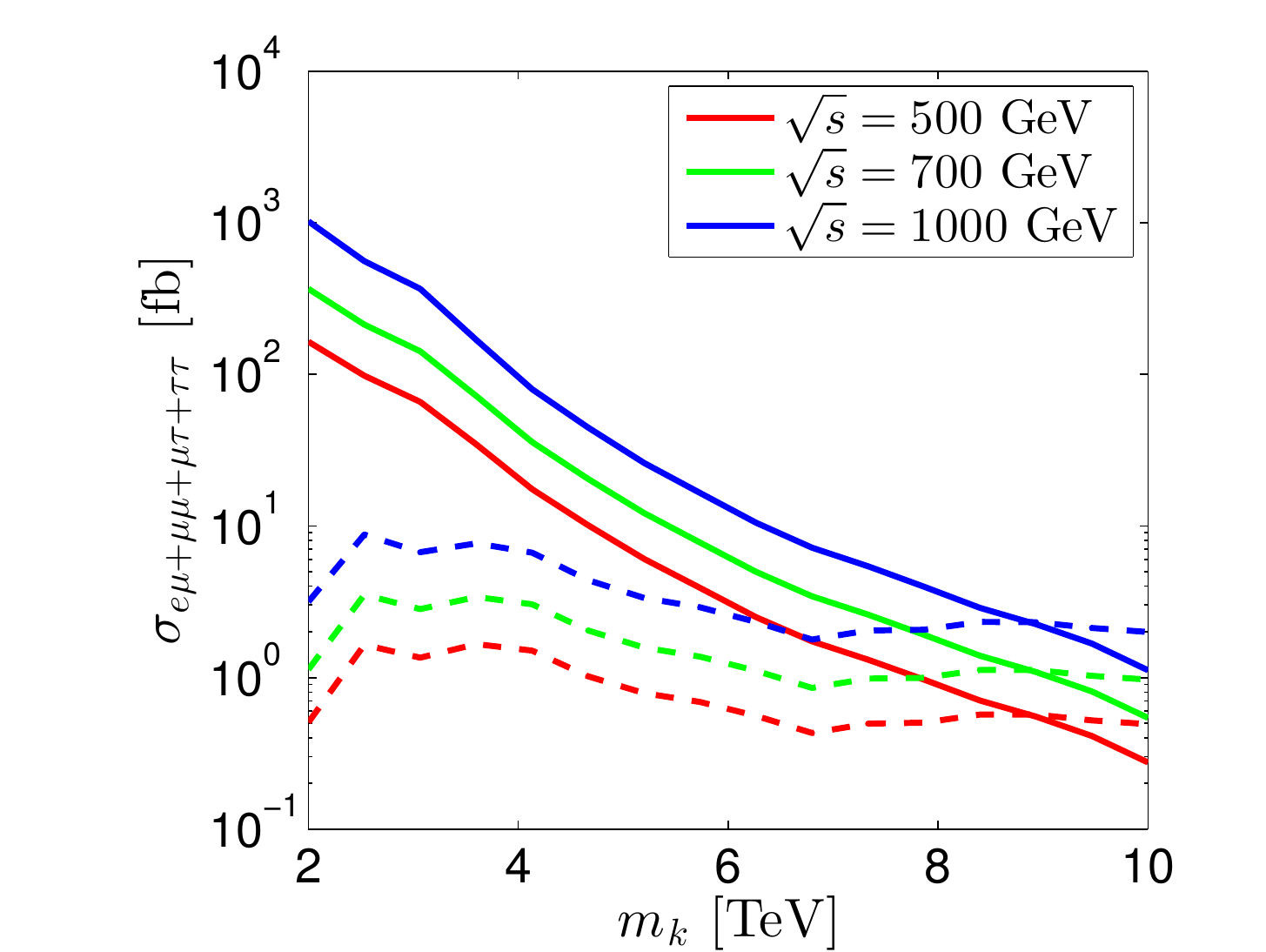}
\includegraphics[width=0.45\textwidth]{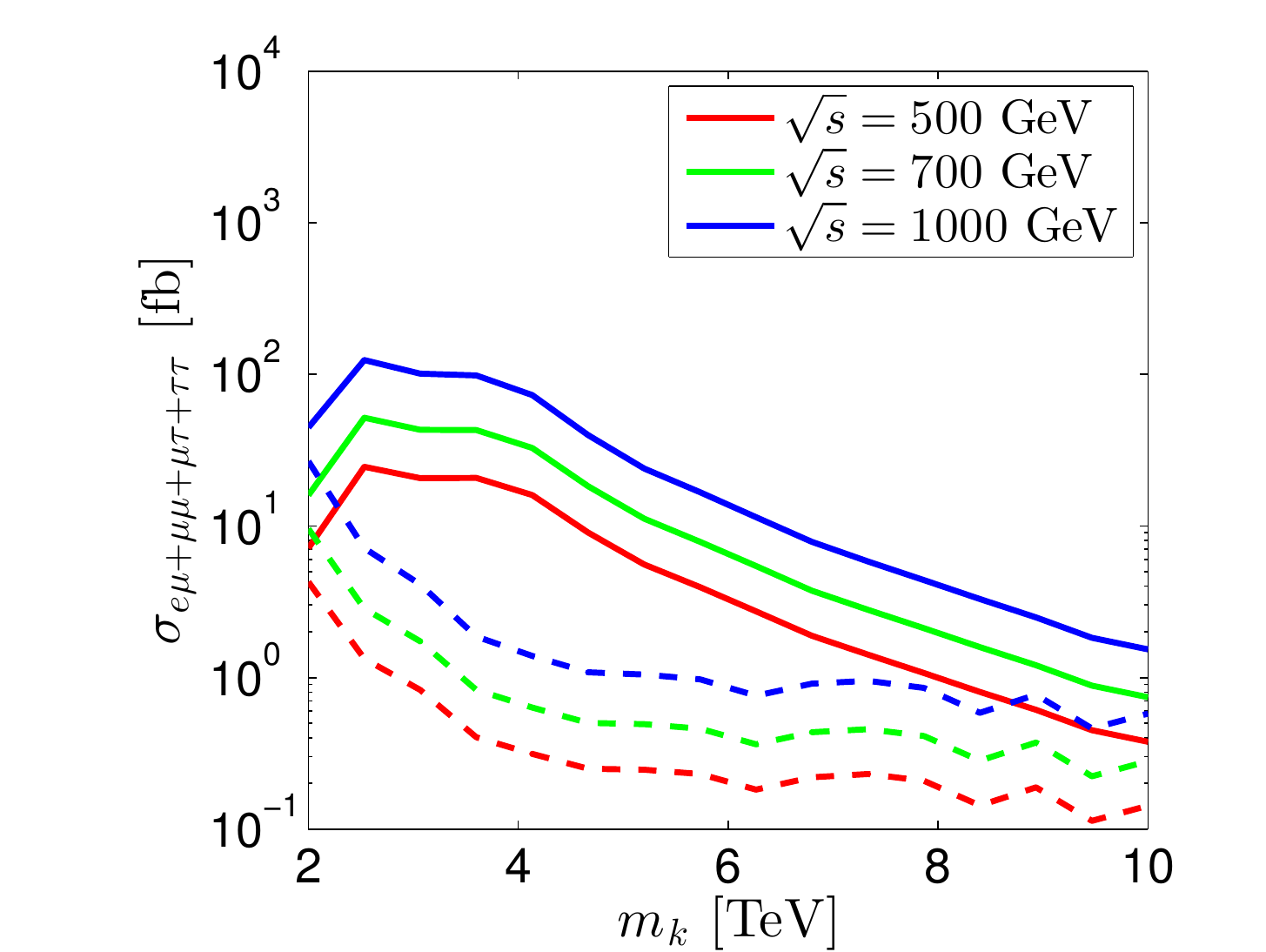}
\caption{\label{fig:ee_mm} Upper limits on the sum of the cross
  sections $\sigma(e^-e^-\to \mu^-\mu^-)+\sigma(e^-e^-\to
  \mu^-e^-)+\sigma(e^-e^-\to \mu^-\tau^-)+\sigma(e^-e^-\to
  \tau^-\tau^-)$ as a function of the
  doubly charged scalar mass for center of mass energies $\sqrt{s} =
  500, 700, 1000$~GeV. Solid curves correspond to the Zee--Babu model,
  where $\sigma(e^-e^-\to \mu^-\mu^-)$ dominates the sum. Dashed
  curves correspond to the Higgs triplet model, where the lightest
  neutrino mass is allowed to vary between 0 and 0.2~eV.  The left
  (right) panel corresponds to normal (inverted) neutrino mass
  ordering.}
\end{center}
\end{figure*}

We show in Fig.~\ref{fig:ee_mm} the upper limits on the cross section as a function of the scalar
mass for center of mass energies $\sqrt{s} = 500, 700, 1000$~GeV (solid curves). For a total
luminosity of 50~${\rm fb}^{-1}$, more than a few tens of events can be expected for a scalar mass
$m_k \lesssim 10~{\rm TeV}$. Also note that, for a smaller $m_k$, the reduction of the cross
section for IO is due to the LFV constraints. While such a signature will not allow for the direct
discovery of the doubly charged scalar via a resonance, it would provide indirect evidence for a
doubly charged particle. The flavour and chirality structure of the LFV processes $e^-e^-\to
\alpha^-\beta^-$ would offer additional consistency checks with the Zee--Babu model, as we are
going to discuss in the next section.

\section{Distinguishing the Zee--Babu and the Higgs triplet model}
\label{sec:type-II}

Once a signal induced by a doubly charged scalar is found at a
collider experiment, an interesting task will be to identify the
underlying model and establish the connection to the mechanism to
generate neutrino mass. Here we focus on ways to distinguish the
Zee--Babu model from the Higgs triplet model, which also predicts the
existence of a doubly charged scalar particle. The leptonic part of
the Lagrangian contains the term
\begin{eqnarray}
{\cal L}_\Delta = h_{\alpha \beta} \overline{L^c}_{\alpha} i\tau_2
\Delta L_\beta + {\rm H.c.},  \label{eq:Ltriplet}
\end{eqnarray}
where $h_{\alpha\beta}$ is a symmetric Yukawa coupling matrix and
$\Delta$ is a $2\times2$ representation of the $SU(2)_L$ Higgs triplet
containing neutral, singly charged, and doubly charged components. Neutrino masses are generated by the vacuum expectation value (VEV) of the
neutral component. Hence the Yukawa couplings
$h_{\alpha\beta}$ are directly proportional to the neutrino mass
matrix. Below we outline a few possibilities to distinguish the
Zee--Babu model from the Higgs triplet model.

\subsection{Signatures at a like-sign linear collider}

Similar as in the Zee--Babu model, also in the Higgs triplet model the process $e^-e^- \to \alpha^-
\beta^-$ is possible at a like-sign linear collider with a cross section in complete analogy to
Eq.~\eqref{eq:XS-likesign} (see e.g.,~\cite{Rodejohann:2010bv}), with the coupling
$g_{\alpha\beta}$ replaced by $h_{\alpha\beta}$ and $S \to (1-P_1)(1-P_2)$, taking into account
that for the triplet left-handed leptons couple to the doubly charged scalar, in contrast to the
right-handed coupling in the Zee--Babu model. In Fig.~\ref{fig:ee_mm} we compare the maximum
obtainable value for the sum of the cross sections for lepton flavour violating processes for the
Higgs triplet model (dashed) to the one for the Zee--Babu model (solid).  For the Higgs triplet
model, the lightest neutrino mass is not necessarily vanishing, therefore, in Fig.~\ref{fig:ee_mm}
we vary its value between zero and 0.2~eV. The VEV of the triplet is varied between 0.1~eV and
1~keV, where the largest cross sections are obtained for small values, since in this case Yukawa
couplings are largest. The two Majorana phases are allowed to vary freely between 0 and $\pi$, and
the Dirac phase $\delta$ between 0 and $2\pi$. As for the Zee--Babu model, we use the current
experimental bounds, i.e., the same constraints as for the black curves in Fig.~\ref{fig:fig1}. For
the triplet model we include the constraints from $\mu\to e\gamma$, $\mu\to 3e$,
muonium--antimuonium conversion, and $\mu - e$ conversion in nuclei. The corresponding expression
can be found for instance in \cite{Ma:2000xh, Kitano:2002mt, Dinh:2012bp}. We find cross sections
of order 1~fb, and those results suggest that with integrated luminosities of $\gtrsim 10$~fb$^{-1}$
such lepton flavour violating processes can also be expected to be observed in the Higgs triplet
model. Note that here we are in the regime of $\sqrt{s}$ much smaller than the mass of the doubly
charged scalar, which implies that no resonance is seen and hence the mass cannot be determined.
Furthermore, we stress again that those curves are upper bounds, and in particular, for the
Zee--Babu model the cross section can be easily reduced by adjusting $g_{ee}$ which is not bounded
from below. Therefore the size of the cross section by itself does not allow to distinguish the two
models.

\begin{figure*}[!t]
\begin{center}
\includegraphics[width=0.45\textwidth]{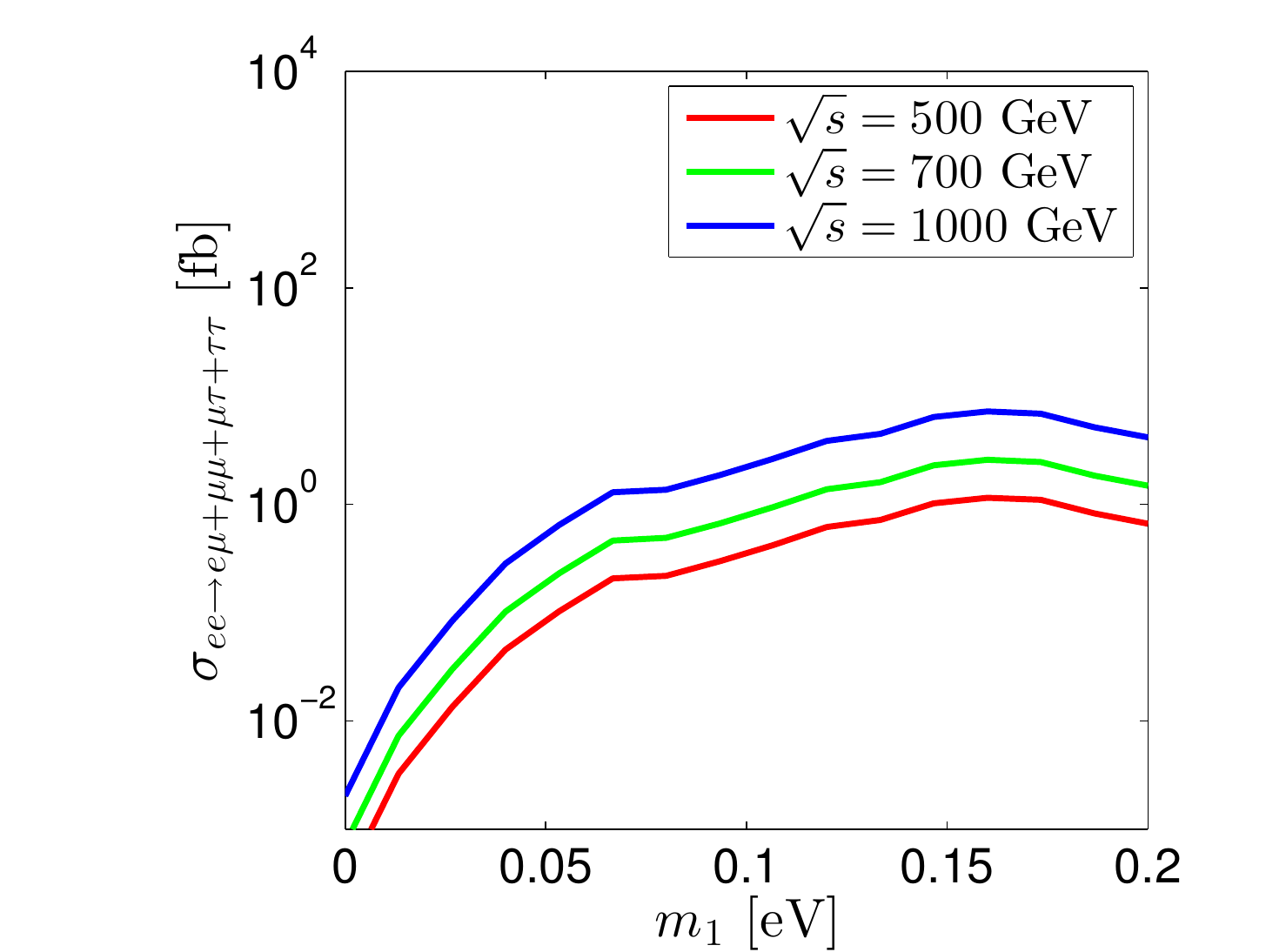}
\includegraphics[width=0.45\textwidth]{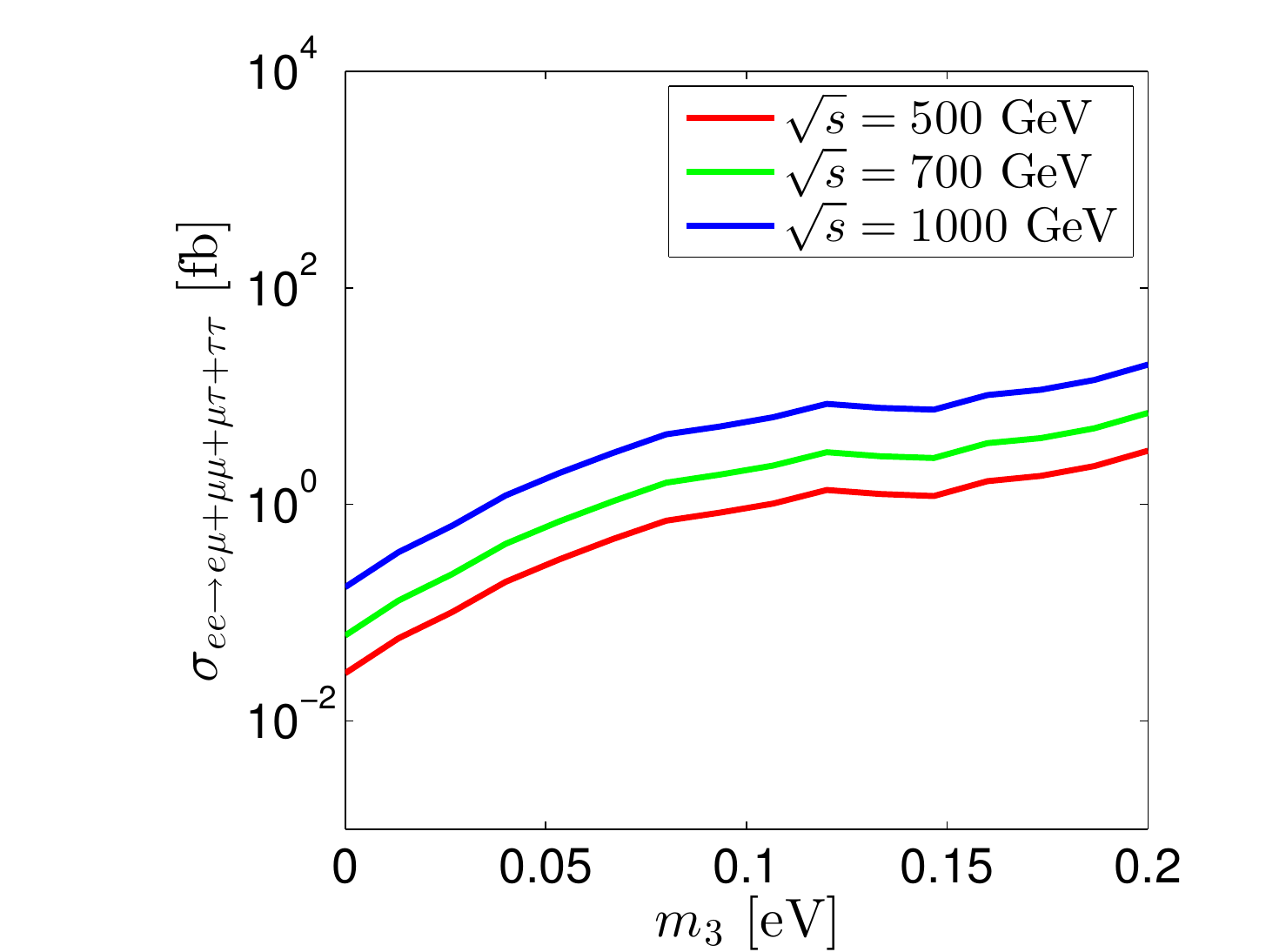}
\caption{\label{fig:XS-HTM} Upper limits on the sum of the cross
  sections $\sigma(e^-e^-\to \mu^-\mu^-)+\sigma(e^-e^-\to
  \mu^-e^-)+\sigma(e^-e^-\to \mu^-\tau^-)+\sigma(e^-e^-\to
  \tau^-\tau^-)$ in the Higgs triplet model as a function of the
  lightest neutrino mass for center of mass energies $\sqrt{s} = 500,
  700, 1000$~GeV. The doubly charged scalar mass is set to 2~TeV.  The
  left (right) panel corresponds to normal (inverted) neutrino mass
  ordering.}
\end{center}
\end{figure*}

Before we discuss the possibility to use the flavour structure to distinguish the models, let us
mention the importance of the lightest neutrino mass in the case of the triplet model.  Since the
cross section for $e^-e^- \to \alpha^- \beta^-$ is proportional to $h_{ee} \propto m^{(\nu)}_{ee}$ the
value of the lightest neutrino mass is important, especially for normal mass ordering, where the
possible size of $m^{(\nu)}_{ee}$ depends strongly on $m_1$. In Fig.~\ref{fig:XS-HTM} we show the maximum
obtainable value for the sum of the cross sections for lepton flavour violating processes as a
function of the lightest neutrino mass. We observe that for strongly hierarchical spectrum with
normal ordering the cross section becomes very small. Hence, establishing normal mass ordering by
oscillation experiments plus setting an upper bound on the lightest neutrino mass below 0.05~eV
(for instance by cosmology or neutrinoless double beta decay) would make the signal in the triplet
model very small. In that case a sizeable signal at the like-sign collider would favour the
Zee--Babu model. Note also that even for large neutrino masses no relevant lower bound on the cross
section can be derived in the triplet model, since Yukawa couplings can be made very small by
increasing the VEV of the triplet above the keV range (up to the GeV scale).

\subsection{Flavour structure of the Yukawa couplings}

Let us now assume that either a leptonically decaying doubly charged resonance is found at LHC or a
linear collider, or that lepton flavour violating processes $e^-e^- \to \alpha^- \beta^-$ are seen
at a like-sign linear collider. In such a case the flavour structure of the decays or the LFV
processes will be rather different in the two cases. Note that ratios of decay rates and LFV cross
sections will be the same and proportional to the corresponding Yukawa couplings:
\begin{align}
R^{\alpha_1\beta_1}_{\alpha_2\beta_2} & \equiv
  \frac{\Gamma(k^{++}\to \alpha_1^+\beta_1^+)}{\Gamma(k^{++}\to \alpha_2^+\beta_2^+)}
=
  \frac{\sigma(e^-e^- \to \alpha_1^-\beta_1^-)}{\sigma(e^-e^-\to \alpha_2^-\beta_2^-)}
\end{align}
where in the Zee--Babu model we have
\begin{equation}
R^{\alpha_1\beta_1}_{\alpha_2\beta_2}
   = \frac{(1 + \delta_{\alpha_1\beta_1}) |g_{\alpha_1\beta_1}|^2}
          {(1 + \delta_{\alpha_2\beta_2}) |g_{\alpha_2\beta_2}|^2} \,,
\end{equation}
and in the triplet model an analogous relation holds but replacing the Yukawa coupling $g$ by $h$.
The important observation is that the flavour structure of those couplings will be rather different
in the two models, with $h_{\alpha\beta}$ proportional to the neutrino mass matrix
$m^{(\nu)}_{\alpha\beta}$, while for $g_{\alpha\beta}$ the relation to the neutrino mass matrix is more
complicated.

\begin{figure*}[!t]
\begin{center}
\includegraphics[width=0.45\textwidth]{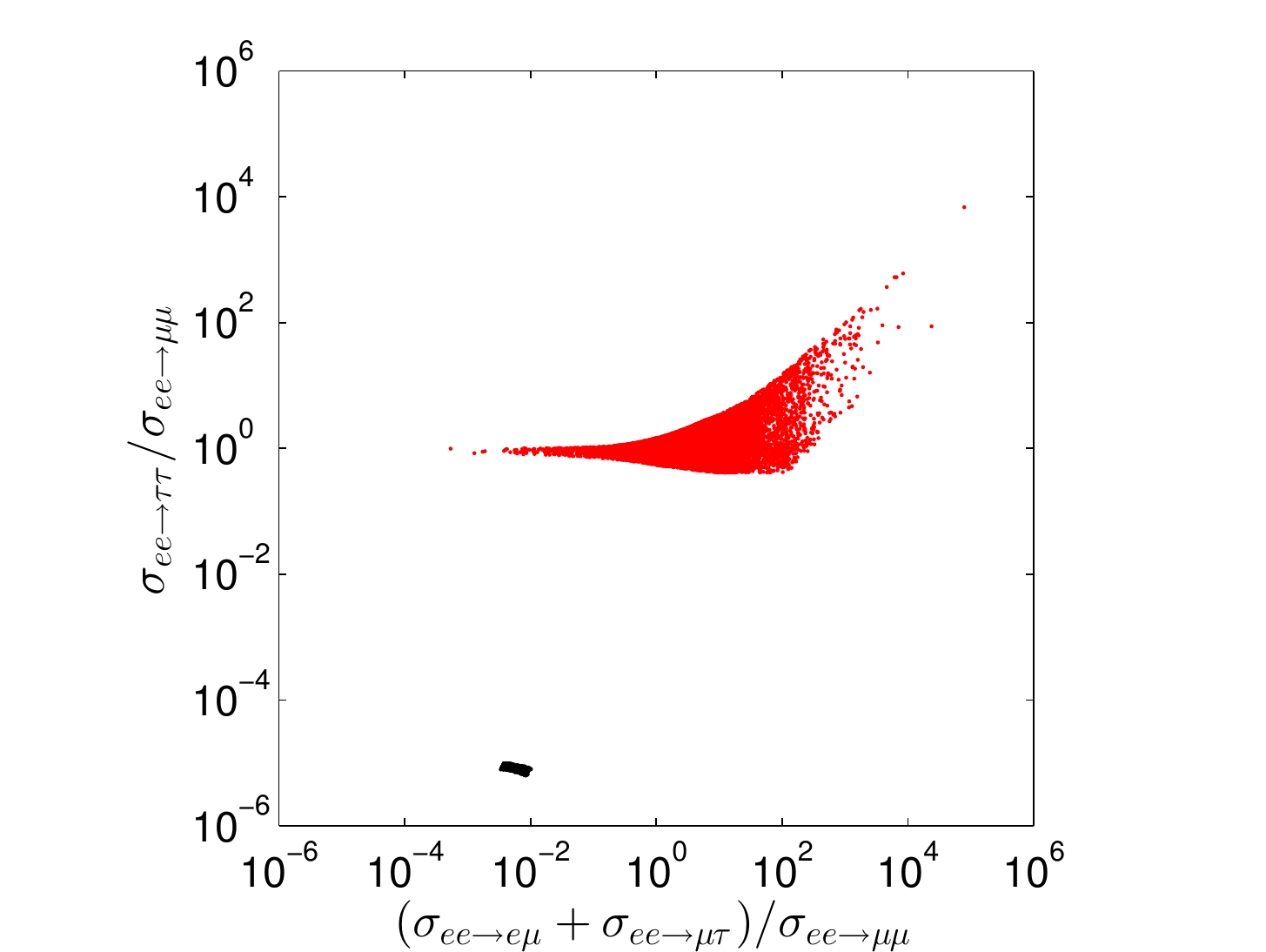}
\includegraphics[width=0.45\textwidth]{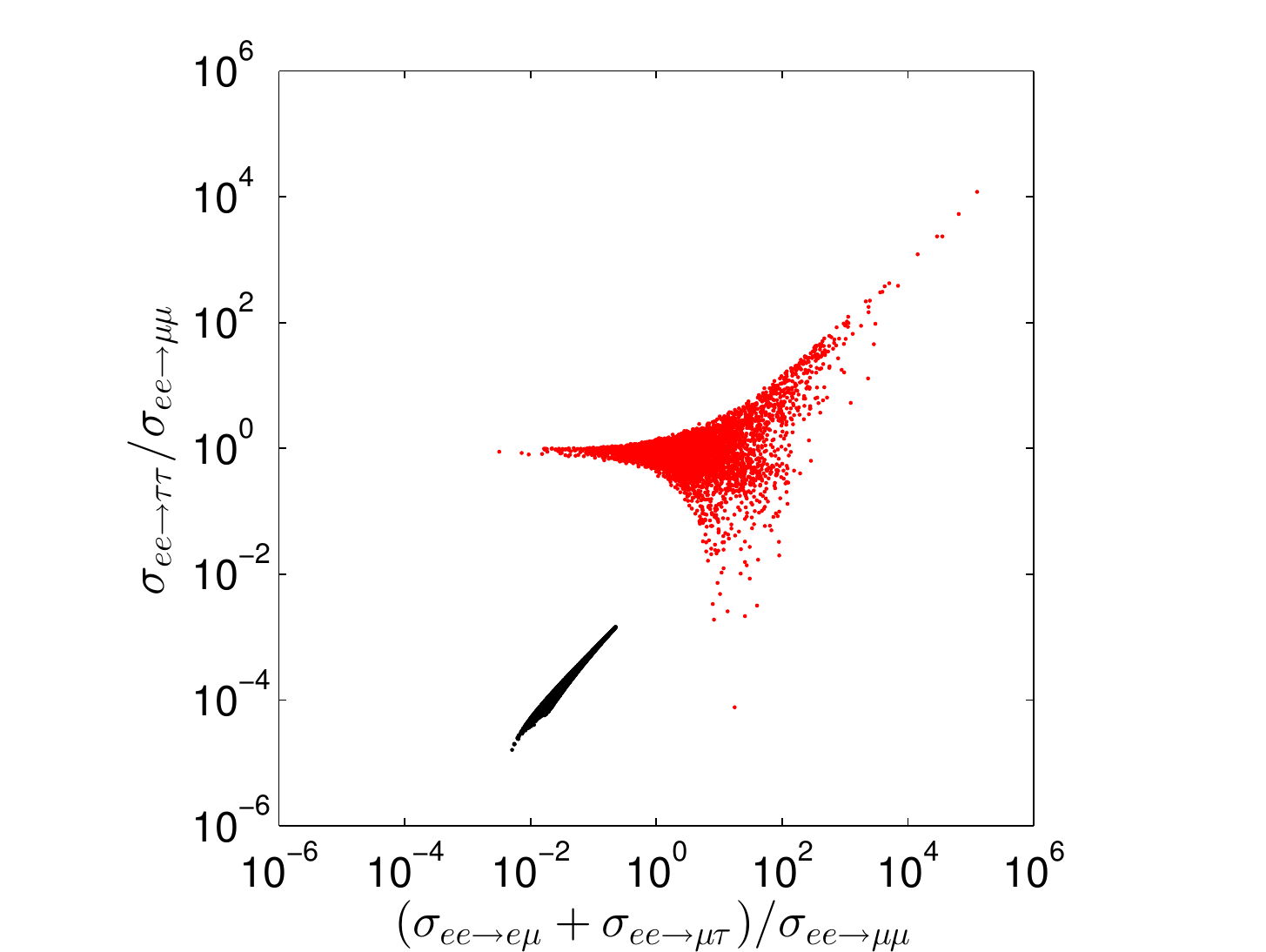}
\caption{\label{fig:cross_ratio} Cross section ratios for processes
  $e^-e^-\to \ell^-_\alpha\ell^-_\beta$ (or equivalently ratios of
  branching fractions of the doubly charged scalar decays). The left
  (right) panel corresponds to normal (inverted) neutrino mass
  ordering. Black points denote the ratios in the Zee--Babu model,
  while red points correspond to the ratios in the Higgs triplet
  model. All scalar masses are taken to be $3~{\rm TeV}$. For the
  triplet model we vary the Majorana phases, the lightest neutrino
  mass between zero and $0.2~{\rm eV}$, and the triplet VEV between
  0.1~eV and 1 keV.}
\end{center}
\end{figure*}

We illustrate in Fig.~\ref{fig:cross_ratio} the cross section ratios
of the LFV processes (or equivalently decay branching fractions) in
the two models.  For the sake of definiteness, all the scalar masses
are taken to be $3~{\rm TeV}$. Those results depend only weakly on the
scalar masses.  The different flavour structure of the two models can
be clearly seen from the plots. In particular, in the Zee--Babu model,
the dominating decay mode is always the $\mu\mu$ channel, no matter of
the neutrino mass ordering. An observation of a significant fraction
of events different from the di-muon channel would exclude the
model. The largest contribution of a different flavour combination may
occur for IO with a fraction of $\mu\tau$ events with
$R^{\mu\tau}_{\mu\mu} \lesssim 0.2$.

In order to see this point more clearly, we insert the neutrino mixing
parameters into Eq.~\eqref{eq:g} and obtain for the NO case
\begin{eqnarray}\label{eq:omegaNH}
\zeta f^2_{\mu\tau} \omega_{\mu\mu} & \simeq m^{(\nu)}_{33} \simeq & m_3 c^2_{23} \; ,\nonumber \\
\zeta f^2_{\mu\tau} \omega_{\mu\tau} & \simeq -m^{(\nu)}_{23} \simeq & -m_3 s_{23} c_{23} \; ,\nonumber \\
\zeta f^2_{\mu\tau} \omega_{\tau\tau} & \simeq m^{(\nu)}_{22} \simeq & m_3 s^2_{23} \; ,
\end{eqnarray}
where $\omega_{\alpha\beta} = g_{\alpha\beta} m_\alpha m_\beta$  (defined after Eq.~\eqref{eq:g}) and
terms proportional to the small parameters $s_{13}$, $m_2$, and $m_e$ have been
neglected. Therefore, one has approximately $|\omega_{\mu\mu}|
\simeq |\omega_{\mu\tau}| \simeq |\omega_{\tau\tau}|$ for a nearly
maximal $\theta_{23}$, and the ratios between Yukawa couplings are
\begin{eqnarray}\label{eq:g-ratios}
g_{\mu\mu}:g_{\mu\tau}:g_{\tau\tau} \sim
1:\frac{m_\mu}{m_\tau}:\frac{m^2_\mu}{m^2_\tau} \; .
\end{eqnarray}
Since the branching ratios are proportional to $|g_{\alpha\beta}|^2$,
one finds $R^{\tau\tau}_{\mu\mu} \simeq m_\mu^4/m_\tau^4 \approx
10^{-5}$ and $R^{\mu\tau}_{\mu\mu} \simeq m_\mu^2/m_\tau^2 \approx
3\times 10^{-3}$, in good agreement with the left plot of
Fig.~\ref{fig:cross_ratio}.  Using the first line in
Eq.~\eqref{eq:omegaNH} with $m_3 = \sqrt{\Delta m^2_{31}} \approx
0.05$~eV, one finds that $g_{\mu\mu}$ is bounded from below by
neutrino masses \cite{Nebot:2007bc}. Numerically we find
\begin{eqnarray}\label{eq:g-bound}
|g_{\mu\mu}||f_{\mu\tau}|^2 \gtrsim 10^{-3}\frac{m_h{\rm
max}(m_k,m_h)}{\mu \, {\rm TeV}} \quad\text{(NO).}
\end{eqnarray}
Taking all the scalar masses to be 3~TeV as an example, the above
condition indicates $|g_{\mu\mu}|\gtrsim 0.3$, where we have used
Eq.~\eqref{eq:f-NO} and the constraint from $\mu\to e \gamma$ on
$f_{\mu\tau}$ from Tab.~\ref{tab:1}. This lower bound on $g_{\mu\mu}$
is in good agreement with our numerical results.

In the case of IO the ratios in Eq.~\eqref{eq:g-ratios} hold only
approximately. Because of the relations in Eq.~\eqref{eq:eqIH3},
cancellations between the various terms in Eq.~\eqref{eq:g} become
possible, leading to the correlation visible in the right panel of
Fig.~\ref{fig:cross_ratio}. We have checked numerically that for IO
the lower bound corresponding to Eq.~\eqref{eq:g-bound} is one order
of magnitude weaker. Numerically we obtain a lower bound of
$|g_{\mu\mu}| \gtrsim 0.1$ for scalar masses of 3~TeV, which can be
understood by using Eq.~\eqref{eq:eqIH3} and the bound on $f_{e\mu}$
from CKM unitarity shown in Tab.~\ref{tab:1}.

Finally, the $g_{ee}$ contribution to the neutrino masses is strongly suppressed by the electron
mass, and thus is allowed to be relatively sizeable compared to other Yukawa couplings. In the case
that $g_{\mu\mu}$ lies close to its lower bound, the dominating decay channel could be $ee$
instead of $\mu\mu$. The most interesting channels are then the $\mu\mu$ and $ee$ channels. Note
however, that this channel is not observable at the like-sign collider due to the Standard Model
M{\o}ller scattering background. Hence this signature can only be explored if decays of an on-shell
doubly charged scalar are observed.

\subsection{Further different signatures}

Apart from exploring the flavour structure of Yukawas there are several
different signatures to distinguish the two models. In particular, for the
triplet model there is large literature on additional
observables. Below we give a brief review of a few possibilities.

Starting with collider signatures, we note that observing a doubly
charged resonance below the lower bounds in the Zee--Babu model
(Eqs.~\eqref{eq:massbound} or \eqref{eq:massbound-5}) would favour the
Higgs triplet. In such a case exploring the flavour structure of the
decays (as illustrated in Fig.~\ref{fig:cross_ratio}) may be used to
establish the relation to neutrino mass, see for instance
\cite{Garayoa:2007fw}.  If a resonance is observed consistent with the
Zee--Babu bounds, one may also look for the singly charged scalar,
which is predicted in both models. In the triplet model the mass
difference between the singly and doubly charged scalars is given by
the VEV of the Higgs doublet, $v$, times a
dimensionless coupling in the scalar potential. Therefore, generically
one expects a mass difference $\lesssim v$. In the Zee--Babu model the
two scalar masses are unrelated. Moreover, the triplet couples to
$W^\pm$, which allows processes like for instance $H^+ \to H^{++}W^-$.
Signatures of the singly charged triplet component have been
investigated in \cite{Perez:2008ha, Akeroyd:2009hb, Aoki:2011pz,
  Akeroyd:2011zza}.

Another difference of the models is that the doubly charged scalar in
the Zee--Babu model (triplet model) couples to right-handed
(left-handed) leptons. In the case of LFV processes at a like-sign
linear collider one can use the polarization of the beams to find out
the chirality structure of the effective operator induced by the heavy
doubly charged scalar~\cite{Cuypers:1997qg}. The possibility to
determine the chirality at a hadron collider by using tau decays has
been investigated in \cite{Sugiyama:2012yw}.

Apart from collider experiments, measurement sensitive to the absolute neutrino mass scale will be
important, see \cite{Weinheimer:2013hya} for a recent review. Since the Zee--Babu model predicts
the lightest neutrino mass to be zero, it can be ruled out by establishing a non-zero lightest
neutrino mass for instance in neutrinoless double beta decay experiments, kinematical neutrino mass
measurement, and/or in cosmology, eventually combined with a determination of the neutrino mass
ordering from oscillation experiments \cite{Blennow:2013oma}.

\section{Conclusions}
\label{sec:summary}

We have studied the current experimental constraints on the Zee--Babu
model, taking into account recent data on lepton mixing angles and the
MEG limit on $\mu\to e \gamma$. By performing a numerical parameter
scan of the model we find that most likely the charged scalars of the
Zee--Babu model will be out of reach for the Large Hadron Collider
(LHC), including the 14~TeV configuration. If a signal should indeed
be seen at LHC this would push the model into a fine tuned parameter
region close to the limit of perturbativity. In such a case a signal
in upcoming experiments searching for charged lepton flavour
violation, such as $\mu\to e$ conversion on nuclei or $\mu\to e
\gamma$ is guaranteed.

Even if the doubly charged scalar of the model is too heavy to be produced at a collider we point
out that a sub-TeV linear collider operated in the like-sign mode may reveal lepton flavour
violating processes $e^-e^- \to \alpha^-\beta^-$ due to contact interactions induced by the heavy
doubly charged scalar. Assuming luminosities of several 10~fb$^{-1}$ such processes might be
observable for scalar masses up to 10~TeV. We stress however, that no signal can be guaranteed,
since it is proportional to the Yukawa coupling $g_{ee}$ which is essentially unconstrained by
neutrino data.

Furthermore we have considered the same signature for an alternative model for neutrino mass, the
Higgs triplet model, which has a similar particle spectrum as the Zee--Babu model, although the
mechanism of neutrino mass generation is very different. We have shown that those two models lead
to a very different flavour structure of LFV signatures at a like-sign collider (or equivalently to
ratios of branching fractions of doubly charged scalar decays, in case they are kinematically
accessible). We have outlined various characteristic signatures of the two models. If neutrino mass
should indeed be generated by one of those two extensions of the scalar sector it seems likely that
the correct model can be identified by using an interplay of various collider signatures as well as
absolute neutrino mass measurement.

\bigskip

{\bf Note:} During the final stage of this work we became aware of
Ref.~\cite{Herrero} where also an updated parameter scan in the
Zee--Babu model is performed. Taking into account the slightly
different method to derive the allowed parameter region and different
perturbativity and fine-tuning requirements our results are
consistent.

\section*{Acknowledgment}

We thank the authors of \cite{Herrero} for sharing their preliminary
results with us.  One of the authors (H.Z.) is indebted to Werner
Rodejohann for useful discussions. This work was supported by the
International Max Planck Research School for Precision Tests of
Fundamental Symmetries (D.S.)  and the Max Planck Society in the
project MANITOP (H.Z.).  T.S.\ acknowledges partial support from the
European Union FP7 ITN INVISIBLES (Marie Curie Actions,
PITN-GA-2011-289442).


\bibliography{bib-babu}

\begin{thebibliography}{58}
\expandafter\ifx\csname natexlab\endcsname\relax\def\natexlab#1{#1}\fi
\expandafter\ifx\csname bibnamefont\endcsname\relax
  \def\bibnamefont#1{#1}\fi
\expandafter\ifx\csname bibfnamefont\endcsname\relax
  \def\bibfnamefont#1{#1}\fi
\expandafter\ifx\csname citenamefont\endcsname\relax
  \def\citenamefont#1{#1}\fi
\expandafter\ifx\csname url\endcsname\relax
  \def\url#1{\texttt{#1}}\fi
\expandafter\ifx\csname urlprefix\endcsname\relax\def\urlprefix{URL }\fi
\providecommand{\bibinfo}[2]{#2}
\providecommand{\eprint}[2][]{\url{#2}}

\bibitem[{\citenamefont{Farzan et~al.}(2013)\citenamefont{Farzan, Pascoli, and
  Schmidt}}]{Farzan:2012ev}
\bibinfo{author}{\bibfnamefont{Y.}~\bibnamefont{Farzan}},
  \bibinfo{author}{\bibfnamefont{S.}~\bibnamefont{Pascoli}}, \bibnamefont{and}
  \bibinfo{author}{\bibfnamefont{M.~A.} \bibnamefont{Schmidt}},
  \bibinfo{journal}{JHEP} \textbf{\bibinfo{volume}{JHEP03}},
  \bibinfo{pages}{107} (\bibinfo{year}{2013}), \eprint{1208.2732}.

\bibitem[{\citenamefont{Angel et~al.}(2013)\citenamefont{Angel, Rodd, and
  Volkas}}]{Angel:2012ug}
\bibinfo{author}{\bibfnamefont{P.~W.} \bibnamefont{Angel}},
  \bibinfo{author}{\bibfnamefont{N.~L.} \bibnamefont{Rodd}}, \bibnamefont{and}
  \bibinfo{author}{\bibfnamefont{R.~R.} \bibnamefont{Volkas}},
  \bibinfo{journal}{Phys. Rev.} \textbf{\bibinfo{volume}{D87}},
  \bibinfo{pages}{073007} (\bibinfo{year}{2013}), \eprint{1212.6111}.

\bibitem[{\citenamefont{Law and McDonald}(2013)}]{Law:2013dya}
\bibinfo{author}{\bibfnamefont{S.~S.~C.} \bibnamefont{Law}} \bibnamefont{and}
  \bibinfo{author}{\bibfnamefont{K.~L.} \bibnamefont{McDonald}}
  (\bibinfo{year}{2013}), \eprint{1303.6384}.

\bibitem[{\citenamefont{Konetschny and Kummer}(1977)}]{Konetschny:1977bn}
\bibinfo{author}{\bibfnamefont{W.}~\bibnamefont{Konetschny}} \bibnamefont{and}
  \bibinfo{author}{\bibfnamefont{W.}~\bibnamefont{Kummer}},
  \bibinfo{journal}{Phys. Lett.} \textbf{\bibinfo{volume}{B70}},
  \bibinfo{pages}{433} (\bibinfo{year}{1977}).

\bibitem[{\citenamefont{Cheng and Li}(1980)}]{Cheng:1980qt}
\bibinfo{author}{\bibfnamefont{T.~P.} \bibnamefont{Cheng}} \bibnamefont{and}
  \bibinfo{author}{\bibfnamefont{L.-F.} \bibnamefont{Li}},
  \bibinfo{journal}{Phys. Rev.} \textbf{\bibinfo{volume}{D22}},
  \bibinfo{pages}{2860} (\bibinfo{year}{1980}).

\bibitem[{\citenamefont{Zee}(1985)}]{Zee:1985rj}
\bibinfo{author}{\bibfnamefont{A.}~\bibnamefont{Zee}}, \bibinfo{journal}{Phys.
  Lett.} \textbf{\bibinfo{volume}{B161}}, \bibinfo{pages}{141}
  (\bibinfo{year}{1985}).

\bibitem[{\citenamefont{Zee}(1986)}]{Zee:1985id}
\bibinfo{author}{\bibfnamefont{A.}~\bibnamefont{Zee}}, \bibinfo{journal}{Nucl.
  Phys.} \textbf{\bibinfo{volume}{B264}}, \bibinfo{pages}{99}
  (\bibinfo{year}{1986}).

\bibitem[{\citenamefont{Babu}(1988)}]{Babu:1988ki}
\bibinfo{author}{\bibfnamefont{K.~S.} \bibnamefont{Babu}},
  \bibinfo{journal}{Phys. Lett.} \textbf{\bibinfo{volume}{B203}},
  \bibinfo{pages}{132} (\bibinfo{year}{1988}).

\bibitem[{\citenamefont{Lindner et~al.}(2011)\citenamefont{Lindner, Schmidt,
  and Schwetz}}]{Lindner:2011it}
\bibinfo{author}{\bibfnamefont{M.}~\bibnamefont{Lindner}},
  \bibinfo{author}{\bibfnamefont{D.}~\bibnamefont{Schmidt}}, \bibnamefont{and}
  \bibinfo{author}{\bibfnamefont{T.}~\bibnamefont{Schwetz}},
  \bibinfo{journal}{Phys. Lett.} \textbf{\bibinfo{volume}{B705}},
  \bibinfo{pages}{324} (\bibinfo{year}{2011}), \eprint{1105.4626}.

\bibitem[{\citenamefont{Baek et~al.}(2012)\citenamefont{Baek, Ko, and
  Senaha}}]{Baek:2012ub}
\bibinfo{author}{\bibfnamefont{S.}~\bibnamefont{Baek}},
  \bibinfo{author}{\bibfnamefont{P.}~\bibnamefont{Ko}}, \bibnamefont{and}
  \bibinfo{author}{\bibfnamefont{E.}~\bibnamefont{Senaha}}
  (\bibinfo{year}{2012}), \eprint{1209.1685}.

\bibitem[{\citenamefont{Babu and Macesanu}(2003)}]{Babu:2002uu}
\bibinfo{author}{\bibfnamefont{K.~S.} \bibnamefont{Babu}} \bibnamefont{and}
  \bibinfo{author}{\bibfnamefont{C.}~\bibnamefont{Macesanu}},
  \bibinfo{journal}{Phys. Rev.} \textbf{\bibinfo{volume}{D67}},
  \bibinfo{pages}{073010} (\bibinfo{year}{2003}), \eprint{hep-ph/0212058}.

\bibitem[{\citenamefont{Aristizabal~Sierra and
  Hirsch}(2006)}]{AristizabalSierra:2006gb}
\bibinfo{author}{\bibfnamefont{D.}~\bibnamefont{Aristizabal~Sierra}}
  \bibnamefont{and} \bibinfo{author}{\bibfnamefont{M.}~\bibnamefont{Hirsch}},
  \bibinfo{journal}{JHEP} \textbf{\bibinfo{volume}{12}}, \bibinfo{pages}{052}
  (\bibinfo{year}{2006}), \eprint{hep-ph/0609307}.

\bibitem[{\citenamefont{Nebot et~al.}(2008)\citenamefont{Nebot, Oliver, Palao,
  and Santamaria}}]{Nebot:2007bc}
\bibinfo{author}{\bibfnamefont{M.}~\bibnamefont{Nebot}},
  \bibinfo{author}{\bibfnamefont{J.~F.} \bibnamefont{Oliver}},
  \bibinfo{author}{\bibfnamefont{D.}~\bibnamefont{Palao}}, \bibnamefont{and}
  \bibinfo{author}{\bibfnamefont{A.}~\bibnamefont{Santamaria}},
  \bibinfo{journal}{Phys. Rev.} \textbf{\bibinfo{volume}{D77}},
  \bibinfo{pages}{093013} (\bibinfo{year}{2008}), \eprint{arXiv:0711.0483}.

\bibitem[{\citenamefont{Ohlsson et~al.}(2009)\citenamefont{Ohlsson, Schwetz,
  and Zhang}}]{Ohlsson:2009vk}
\bibinfo{author}{\bibfnamefont{T.}~\bibnamefont{Ohlsson}},
  \bibinfo{author}{\bibfnamefont{T.}~\bibnamefont{Schwetz}}, \bibnamefont{and}
  \bibinfo{author}{\bibfnamefont{H.}~\bibnamefont{Zhang}},
  \bibinfo{journal}{Phys. Lett.} \textbf{\bibinfo{volume}{B681}},
  \bibinfo{pages}{269} (\bibinfo{year}{2009}), \eprint{0909.0455}.

\bibitem[{\citenamefont{Abe et~al.}(2012)}]{Abe:2011fz}
\bibinfo{author}{\bibfnamefont{Y.}~\bibnamefont{Abe}} \bibnamefont{et~al.}
  (\bibinfo{collaboration}{DOUBLE-CHOOZ Collaboration}),
  \bibinfo{journal}{Phys. Rev. Lett.} \textbf{\bibinfo{volume}{108}},
  \bibinfo{pages}{131801} (\bibinfo{year}{2012}), \eprint{1112.6353}.

\bibitem[{\citenamefont{An et~al.}(2012)}]{An:2012eh}
\bibinfo{author}{\bibfnamefont{F.}~\bibnamefont{An}} \bibnamefont{et~al.}
  (\bibinfo{collaboration}{DAYA-BAY Collaboration}), \bibinfo{journal}{Phys.
  Rev. Lett.} \textbf{\bibinfo{volume}{108}}, \bibinfo{pages}{171803}
  (\bibinfo{year}{2012}), \eprint{1203.1669}.

\bibitem[{\citenamefont{Ahn et~al.}(2012)}]{Ahn:2012nd}
\bibinfo{author}{\bibfnamefont{J.}~\bibnamefont{Ahn}} \bibnamefont{et~al.}
  (\bibinfo{collaboration}{RENO collaboration}), \bibinfo{journal}{Phys. Rev.
  Lett.} \textbf{\bibinfo{volume}{108}}, \bibinfo{pages}{191802}
  (\bibinfo{year}{2012}), \eprint{1204.0626}.

\bibitem[{\citenamefont{Gonzalez-Garcia
  et~al.}(2012)\citenamefont{Gonzalez-Garcia, Maltoni, Salvado, and
  Schwetz}}]{GonzalezGarcia:2012sz}
\bibinfo{author}{\bibfnamefont{M.}~\bibnamefont{Gonzalez-Garcia}},
  \bibinfo{author}{\bibfnamefont{M.}~\bibnamefont{Maltoni}},
  \bibinfo{author}{\bibfnamefont{J.}~\bibnamefont{Salvado}}, \bibnamefont{and}
  \bibinfo{author}{\bibfnamefont{T.}~\bibnamefont{Schwetz}},
  \bibinfo{journal}{JHEP} \textbf{\bibinfo{volume}{1212}}, \bibinfo{pages}{123}
  (\bibinfo{year}{2012}), \bibinfo{note}{{\tt www.nu-fit.org}},
  \eprint{1209.3023}.

\bibitem[{\citenamefont{Adam et~al.}(2013)}]{Adam:2013mnn}
\bibinfo{author}{\bibfnamefont{J.}~\bibnamefont{Adam}} \bibnamefont{et~al.}
  (\bibinfo{collaboration}{MEG Collaboration}) (\bibinfo{year}{2013}),
  \eprint{1303.0754}.

\bibitem[{\citenamefont{Schechter and Valle}(1980)}]{Schechter:1980gr}
\bibinfo{author}{\bibfnamefont{J.}~\bibnamefont{Schechter}} \bibnamefont{and}
  \bibinfo{author}{\bibfnamefont{J.~W.~F.} \bibnamefont{Valle}},
  \bibinfo{journal}{Phys. Rev.} \textbf{\bibinfo{volume}{D22}},
  \bibinfo{pages}{2227} (\bibinfo{year}{1980}).

\bibitem[{\citenamefont{Lazarides et~al.}(1981)\citenamefont{Lazarides, Shafi,
  and Wetterich}}]{Lazarides:1980nt}
\bibinfo{author}{\bibfnamefont{G.}~\bibnamefont{Lazarides}},
  \bibinfo{author}{\bibfnamefont{Q.}~\bibnamefont{Shafi}}, \bibnamefont{and}
  \bibinfo{author}{\bibfnamefont{C.}~\bibnamefont{Wetterich}},
  \bibinfo{journal}{Nucl. Phys.} \textbf{\bibinfo{volume}{B181}},
  \bibinfo{pages}{287} (\bibinfo{year}{1981}).

\bibitem[{\citenamefont{Mohapatra and Senjanovi{\'c}}(1981)}]{Mohapatra:1980yp}
\bibinfo{author}{\bibfnamefont{R.~N.} \bibnamefont{Mohapatra}}
  \bibnamefont{and}
  \bibinfo{author}{\bibfnamefont{G.}~\bibnamefont{Senjanovi{\'c}}},
  \bibinfo{journal}{Phys. Rev.} \textbf{\bibinfo{volume}{D23}},
  \bibinfo{pages}{165} (\bibinfo{year}{1981}).

\bibitem[{\citenamefont{Chang et~al.}(1988)\citenamefont{Chang, Keung, and
  Pal}}]{Chang:1988aa}
\bibinfo{author}{\bibfnamefont{D.}~\bibnamefont{Chang}},
  \bibinfo{author}{\bibfnamefont{W.-Y.} \bibnamefont{Keung}}, \bibnamefont{and}
  \bibinfo{author}{\bibfnamefont{P.}~\bibnamefont{Pal}},
  \bibinfo{journal}{Phys. Rev. Lett.} \textbf{\bibinfo{volume}{61}},
  \bibinfo{pages}{2420} (\bibinfo{year}{1988}).

\bibitem[{\citenamefont{McDonald and McKellar}(2003)}]{McDonald:2003zj}
\bibinfo{author}{\bibfnamefont{K.~L.} \bibnamefont{McDonald}} \bibnamefont{and}
  \bibinfo{author}{\bibfnamefont{B.~H.~J.} \bibnamefont{McKellar}}
  (\bibinfo{year}{2003}), \eprint{hep-ph/0309270}.

\bibitem[{\citenamefont{Kitano et~al.}(2002)\citenamefont{Kitano, Koike, and
  Okada}}]{Kitano:2002mt}
\bibinfo{author}{\bibfnamefont{R.}~\bibnamefont{Kitano}},
  \bibinfo{author}{\bibfnamefont{M.}~\bibnamefont{Koike}}, \bibnamefont{and}
  \bibinfo{author}{\bibfnamefont{Y.}~\bibnamefont{Okada}},
  \bibinfo{journal}{Phys. Rev.} \textbf{\bibinfo{volume}{D66}},
  \bibinfo{pages}{096002} (\bibinfo{year}{2002}), \eprint{hep-ph/0203110}.

\bibitem[{\citenamefont{Dinh et~al.}(2012)\citenamefont{Dinh, Ibarra, Molinaro,
  and Petcov}}]{Dinh:2012bp}
\bibinfo{author}{\bibfnamefont{D.}~\bibnamefont{Dinh}},
  \bibinfo{author}{\bibfnamefont{A.}~\bibnamefont{Ibarra}},
  \bibinfo{author}{\bibfnamefont{E.}~\bibnamefont{Molinaro}}, \bibnamefont{and}
  \bibinfo{author}{\bibfnamefont{S.}~\bibnamefont{Petcov}},
  \bibinfo{journal}{JHEP} \textbf{\bibinfo{volume}{1208}}, \bibinfo{pages}{125}
  (\bibinfo{year}{2012}), \eprint{1205.4671}.

\bibitem[{\citenamefont{Ma et~al.}(2001)\citenamefont{Ma, Raidal, and
  Sarkar}}]{Ma:2000xh}
\bibinfo{author}{\bibfnamefont{E.}~\bibnamefont{Ma}},
  \bibinfo{author}{\bibfnamefont{M.}~\bibnamefont{Raidal}}, \bibnamefont{and}
  \bibinfo{author}{\bibfnamefont{U.}~\bibnamefont{Sarkar}},
  \bibinfo{journal}{Nucl. Phys.} \textbf{\bibinfo{volume}{B615}},
  \bibinfo{pages}{313} (\bibinfo{year}{2001}), \eprint{hep-ph/0012101}.

\bibitem[{\citenamefont{Beringer et~al.}(2012)}]{Beringer:1900zz}
\bibinfo{author}{\bibfnamefont{J.}~\bibnamefont{Beringer}} \bibnamefont{et~al.}
  (\bibinfo{collaboration}{Particle Data Group}), \bibinfo{journal}{Phys. Rev.}
  \textbf{\bibinfo{volume}{D86}}, \bibinfo{pages}{010001}
  (\bibinfo{year}{2012}).

\bibitem[{\citenamefont{Bennett et~al.}(2006)}]{Bennett:2006fi}
\bibinfo{author}{\bibfnamefont{G.}~\bibnamefont{Bennett}} \bibnamefont{et~al.}
  (\bibinfo{collaboration}{Muon G-2 Collaboration}), \bibinfo{journal}{Phys.
  Rev.} \textbf{\bibinfo{volume}{D73}}, \bibinfo{pages}{072003}
  (\bibinfo{year}{2006}), \eprint{hep-ex/0602035}.

\bibitem[{\citenamefont{Bellgardt et~al.}(1988)}]{Bellgardt:1987du}
\bibinfo{author}{\bibfnamefont{U.}~\bibnamefont{Bellgardt}}
  \bibnamefont{et~al.} (\bibinfo{collaboration}{SINDRUM Collaboration}),
  \bibinfo{journal}{Nucl. Phys.} \textbf{\bibinfo{volume}{B299}},
  \bibinfo{pages}{1} (\bibinfo{year}{1988}).

\bibitem[{\citenamefont{Hayasaka et~al.}(2010)\citenamefont{Hayasaka, Inami,
  Miyazaki, Arinstein, Aulchenko et~al.}}]{Hayasaka:2010np}
\bibinfo{author}{\bibfnamefont{K.}~\bibnamefont{Hayasaka}},
  \bibinfo{author}{\bibfnamefont{K.}~\bibnamefont{Inami}},
  \bibinfo{author}{\bibfnamefont{Y.}~\bibnamefont{Miyazaki}},
  \bibinfo{author}{\bibfnamefont{K.}~\bibnamefont{Arinstein}},
  \bibinfo{author}{\bibfnamefont{V.}~\bibnamefont{Aulchenko}},
  \bibnamefont{et~al.} (\bibinfo{collaboration}{Belle Collaboration}),
  \bibinfo{journal}{Phys. Lett.} \textbf{\bibinfo{volume}{B687}},
  \bibinfo{pages}{139} (\bibinfo{year}{2010}), \eprint{1001.3221}.

\bibitem[{\citenamefont{Aubert et~al.}(2010)}]{Aubert:2009ag}
\bibinfo{author}{\bibfnamefont{B.}~\bibnamefont{Aubert}} \bibnamefont{et~al.}
  (\bibinfo{collaboration}{BaBar Collaboration}), \bibinfo{journal}{Phys. Rev.
  Lett.} \textbf{\bibinfo{volume}{104}}, \bibinfo{pages}{021802}
  (\bibinfo{year}{2010}), \eprint{0908.2381}.

\bibitem[{\citenamefont{Bertl et~al.}(2006)}]{Bertl:2006up}
\bibinfo{author}{\bibfnamefont{W.~H.} \bibnamefont{Bertl}} \bibnamefont{et~al.}
  (\bibinfo{collaboration}{SINDRUM II Collaboration}), \bibinfo{journal}{Eur.
  Phys. J.} \textbf{\bibinfo{volume}{C47}}, \bibinfo{pages}{337}
  (\bibinfo{year}{2006}).

\bibitem[{\citenamefont{Abrams et~al.}(2012)}]{Abrams:2012er}
\bibinfo{author}{\bibfnamefont{R.}~\bibnamefont{Abrams}} \bibnamefont{et~al.}
  (\bibinfo{collaboration}{Mu2e Collaboration}) (\bibinfo{year}{2012}),
  \eprint{1211.7019}.

\bibitem[{\citenamefont{Knoepfel et~al.}(2013)}]{Knoepfel:2013nqa}
\bibinfo{author}{\bibfnamefont{K.}~\bibnamefont{Knoepfel}} \bibnamefont{et~al.}
  (\bibinfo{collaboration}{R. C. Group}) (\bibinfo{year}{2013}),
  \eprint{1307.1168}.

\bibitem[{\citenamefont{Cui et~al.}(2009)}]{Cui:2009zz}
\bibinfo{author}{\bibfnamefont{Y.}~\bibnamefont{Cui}} \bibnamefont{et~al.}
  (\bibinfo{collaboration}{COMET Collaboration}) (\bibinfo{year}{2009}),
  \eprint{KEK-2009-10}.

\bibitem[{\citenamefont{Witte et~al.}(2012)\citenamefont{Witte, Muratori, Hock,
  Appleby, Owen et~al.}}]{Witte:2012zza}
\bibinfo{author}{\bibfnamefont{H.}~\bibnamefont{Witte}},
  \bibinfo{author}{\bibfnamefont{B.}~\bibnamefont{Muratori}},
  \bibinfo{author}{\bibfnamefont{K.}~\bibnamefont{Hock}},
  \bibinfo{author}{\bibfnamefont{R.}~\bibnamefont{Appleby}},
  \bibinfo{author}{\bibfnamefont{H.}~\bibnamefont{Owen}}, \bibnamefont{et~al.},
  \bibinfo{journal}{Conf. Proc.} \textbf{\bibinfo{volume}{C1205201}},
  \bibinfo{pages}{79} (\bibinfo{year}{2012}).

\bibitem[{\citenamefont{del Aguila et~al.}(2013)\citenamefont{del Aguila,
  Chala, Santamaria, and Wudka}}]{delAguila:2013yaa}
\bibinfo{author}{\bibfnamefont{F.}~\bibnamefont{del Aguila}},
  \bibinfo{author}{\bibfnamefont{M.}~\bibnamefont{Chala}},
  \bibinfo{author}{\bibfnamefont{A.}~\bibnamefont{Santamaria}},
  \bibnamefont{and} \bibinfo{author}{\bibfnamefont{J.}~\bibnamefont{Wudka}},
  \bibinfo{journal}{Phys. Lett.} \textbf{\bibinfo{volume}{B725}},
  \bibinfo{pages}{310} (\bibinfo{year}{2013}), \eprint{1305.3904}.

\bibitem[{\citenamefont{Alloul et~al.}(2013)\citenamefont{Alloul, Frank, Fuks,
  and de~Traubenberg}}]{Alloul:2013raa}
\bibinfo{author}{\bibfnamefont{A.}~\bibnamefont{Alloul}},
  \bibinfo{author}{\bibfnamefont{M.}~\bibnamefont{Frank}},
  \bibinfo{author}{\bibfnamefont{B.}~\bibnamefont{Fuks}}, \bibnamefont{and}
  \bibinfo{author}{\bibfnamefont{M.~R.} \bibnamefont{de~Traubenberg}},
  \bibinfo{journal}{Phys. Rev.} \textbf{\bibinfo{volume}{D88}},
  \bibinfo{pages}{075004} (\bibinfo{year}{2013}), \eprint{1307.1711}.

\bibitem[{\citenamefont{Aad et~al.}(2012)}]{ATLAS:2012hi}
\bibinfo{author}{\bibfnamefont{G.}~\bibnamefont{Aad}} \bibnamefont{et~al.}
  (\bibinfo{collaboration}{ATLAS Collaboration}), \bibinfo{journal}{Eur. Phys.
  J.} \textbf{\bibinfo{volume}{C72}}, \bibinfo{pages}{2244}
  (\bibinfo{year}{2012}), \eprint{1210.5070}.

\bibitem[{\citenamefont{Chatrchyan et~al.}(2012)}]{Chatrchyan:2012ya}
\bibinfo{author}{\bibfnamefont{S.}~\bibnamefont{Chatrchyan}}
  \bibnamefont{et~al.} (\bibinfo{collaboration}{CMS Collaboration}),
  \bibinfo{journal}{Eur. Phys. J.} \textbf{\bibinfo{volume}{C72}},
  \bibinfo{pages}{2189} (\bibinfo{year}{2012}), \eprint{1207.2666}.

\bibitem[{\citenamefont{Rizzo}(1982{\natexlab{a}})}]{Rizzo:1981xx}
\bibinfo{author}{\bibfnamefont{T.~G.} \bibnamefont{Rizzo}},
  \bibinfo{journal}{Phys. Rev.} \textbf{\bibinfo{volume}{D25}},
  \bibinfo{pages}{1355} (\bibinfo{year}{1982}{\natexlab{a}}).

\bibitem[{\citenamefont{Rizzo}(1982{\natexlab{b}})}]{Rizzo:1982kn}
\bibinfo{author}{\bibfnamefont{T.~G.} \bibnamefont{Rizzo}},
  \bibinfo{journal}{Phys. Lett.} \textbf{\bibinfo{volume}{B116}},
  \bibinfo{pages}{23} (\bibinfo{year}{1982}{\natexlab{b}}).

\bibitem[{\citenamefont{London et~al.}(1987)\citenamefont{London, Belanger, and
  Ng}}]{London:1987nz}
\bibinfo{author}{\bibfnamefont{D.}~\bibnamefont{London}},
  \bibinfo{author}{\bibfnamefont{G.}~\bibnamefont{Belanger}}, \bibnamefont{and}
  \bibinfo{author}{\bibfnamefont{J.}~\bibnamefont{Ng}}, \bibinfo{journal}{Phys.
  Lett.} \textbf{\bibinfo{volume}{B188}}, \bibinfo{pages}{155}
  (\bibinfo{year}{1987}).

\bibitem[{\citenamefont{Cuypers and Raidal}(1997)}]{Cuypers:1997qg}
\bibinfo{author}{\bibfnamefont{F.}~\bibnamefont{Cuypers}} \bibnamefont{and}
  \bibinfo{author}{\bibfnamefont{M.}~\bibnamefont{Raidal}},
  \bibinfo{journal}{Nucl. Phys.} \textbf{\bibinfo{volume}{B501}},
  \bibinfo{pages}{3} (\bibinfo{year}{1997}), \eprint{hep-ph/9704224}.

\bibitem[{\citenamefont{Raidal}(1998)}]{Raidal:1997tb}
\bibinfo{author}{\bibfnamefont{M.}~\bibnamefont{Raidal}},
  \bibinfo{journal}{Phys. Rev.} \textbf{\bibinfo{volume}{D57}},
  \bibinfo{pages}{2013} (\bibinfo{year}{1998}), \eprint{hep-ph/9706279}.

\bibitem[{\citenamefont{Cakir}(2006)}]{Cakir:2006pa}
\bibinfo{author}{\bibfnamefont{O.}~\bibnamefont{Cakir}}, \bibinfo{journal}{New
  J. Phys.} \textbf{\bibinfo{volume}{8}}, \bibinfo{pages}{145}
  (\bibinfo{year}{2006}), \eprint{hep-ph/0604183}.

\bibitem[{\citenamefont{Rodejohann}(2010)}]{Rodejohann:2010jh}
\bibinfo{author}{\bibfnamefont{W.}~\bibnamefont{Rodejohann}},
  \bibinfo{journal}{Phys. Rev.} \textbf{\bibinfo{volume}{D81}},
  \bibinfo{pages}{114001} (\bibinfo{year}{2010}), \eprint{1005.2854}.

\bibitem[{\citenamefont{Rodejohann and Zhang}(2011)}]{Rodejohann:2010bv}
\bibinfo{author}{\bibfnamefont{W.}~\bibnamefont{Rodejohann}} \bibnamefont{and}
  \bibinfo{author}{\bibfnamefont{H.}~\bibnamefont{Zhang}},
  \bibinfo{journal}{Phys. Rev.} \textbf{\bibinfo{volume}{D83}},
  \bibinfo{pages}{073005} (\bibinfo{year}{2011}), \eprint{1011.3606}.

\bibitem[{\citenamefont{Garayoa and Schwetz}(2008)}]{Garayoa:2007fw}
\bibinfo{author}{\bibfnamefont{J.}~\bibnamefont{Garayoa}} \bibnamefont{and}
  \bibinfo{author}{\bibfnamefont{T.}~\bibnamefont{Schwetz}},
  \bibinfo{journal}{JHEP} \textbf{\bibinfo{volume}{0803}}, \bibinfo{pages}{009}
  (\bibinfo{year}{2008}), \eprint{0712.1453}.

\bibitem[{\citenamefont{Fileviez~Perez
  et~al.}(2008)\citenamefont{Fileviez~Perez, Han, Huang, Li, and
  Wang}}]{Perez:2008ha}
\bibinfo{author}{\bibfnamefont{P.}~\bibnamefont{Fileviez~Perez}},
  \bibinfo{author}{\bibfnamefont{T.}~\bibnamefont{Han}},
  \bibinfo{author}{\bibfnamefont{G.-y.} \bibnamefont{Huang}},
  \bibinfo{author}{\bibfnamefont{T.}~\bibnamefont{Li}}, \bibnamefont{and}
  \bibinfo{author}{\bibfnamefont{K.}~\bibnamefont{Wang}},
  \bibinfo{journal}{Phys. Rev.} \textbf{\bibinfo{volume}{D78}},
  \bibinfo{pages}{015018} (\bibinfo{year}{2008}), \eprint{0805.3536}.

\bibitem[{\citenamefont{Akeroyd and Chiang}(2009)}]{Akeroyd:2009hb}
\bibinfo{author}{\bibfnamefont{A.}~\bibnamefont{Akeroyd}} \bibnamefont{and}
  \bibinfo{author}{\bibfnamefont{C.-W.} \bibnamefont{Chiang}},
  \bibinfo{journal}{Phys. Rev.} \textbf{\bibinfo{volume}{D80}},
  \bibinfo{pages}{113010} (\bibinfo{year}{2009}), \eprint{0909.4419}.

\bibitem[{\citenamefont{Aoki et~al.}(2012)\citenamefont{Aoki, Kanemura, and
  Yagyu}}]{Aoki:2011pz}
\bibinfo{author}{\bibfnamefont{M.}~\bibnamefont{Aoki}},
  \bibinfo{author}{\bibfnamefont{S.}~\bibnamefont{Kanemura}}, \bibnamefont{and}
  \bibinfo{author}{\bibfnamefont{K.}~\bibnamefont{Yagyu}},
  \bibinfo{journal}{Phys. Rev.} \textbf{\bibinfo{volume}{D85}},
  \bibinfo{pages}{055007} (\bibinfo{year}{2012}), \eprint{1110.4625}.

\bibitem[{\citenamefont{Akeroyd and Sugiyama}(2011)}]{Akeroyd:2011zza}
\bibinfo{author}{\bibfnamefont{A.}~\bibnamefont{Akeroyd}} \bibnamefont{and}
  \bibinfo{author}{\bibfnamefont{H.}~\bibnamefont{Sugiyama}},
  \bibinfo{journal}{Phys. Rev.} \textbf{\bibinfo{volume}{D84}},
  \bibinfo{pages}{035010} (\bibinfo{year}{2011}), \eprint{1105.2209}.

\bibitem[{\citenamefont{Sugiyama et~al.}(2012)\citenamefont{Sugiyama, Tsumura,
  and Yokoya}}]{Sugiyama:2012yw}
\bibinfo{author}{\bibfnamefont{H.}~\bibnamefont{Sugiyama}},
  \bibinfo{author}{\bibfnamefont{K.}~\bibnamefont{Tsumura}}, \bibnamefont{and}
  \bibinfo{author}{\bibfnamefont{H.}~\bibnamefont{Yokoya}},
  \bibinfo{journal}{Phys. Lett.} \textbf{\bibinfo{volume}{B717}},
  \bibinfo{pages}{229} (\bibinfo{year}{2012}), \eprint{1207.0179}.

\bibitem[{\citenamefont{Weinheimer and Zuber}(2013)}]{Weinheimer:2013hya}
\bibinfo{author}{\bibfnamefont{C.}~\bibnamefont{Weinheimer}} \bibnamefont{and}
  \bibinfo{author}{\bibfnamefont{K.}~\bibnamefont{Zuber}},
  \bibinfo{journal}{Annalen der Physik,} \textbf{\bibinfo{volume}{525}},
  \bibinfo{pages}{565} (\bibinfo{year}{2013}), \eprint{1307.3518}.

\bibitem[{\citenamefont{Blennow et~al.}(2013)\citenamefont{Blennow, Coloma,
  Huber, and Schwetz}}]{Blennow:2013oma}
\bibinfo{author}{\bibfnamefont{M.}~\bibnamefont{Blennow}},
  \bibinfo{author}{\bibfnamefont{P.}~\bibnamefont{Coloma}},
  \bibinfo{author}{\bibfnamefont{P.}~\bibnamefont{Huber}}, \bibnamefont{and}
  \bibinfo{author}{\bibfnamefont{T.}~\bibnamefont{Schwetz}}
  (\bibinfo{year}{2013}), \eprint{1311.1822}.

\bibitem[{\citenamefont{Herrero-Garcia
  et~al.}(2014)\citenamefont{Herrero-Garcia, Nebot, Rius, and
  Santamaria}}]{Herrero}
\bibinfo{author}{\bibfnamefont{J.}~\bibnamefont{Herrero-Garcia}},
  \bibinfo{author}{\bibfnamefont{M.}~\bibnamefont{Nebot}},
  \bibinfo{author}{\bibfnamefont{N.}~\bibnamefont{Rius}}, \bibnamefont{and}
  \bibinfo{author}{\bibfnamefont{A.}~\bibnamefont{Santamaria}}
  (\bibinfo{year}{2014}), \bibinfo{note}{to appear}.

\end{thebibliography}

\end{document}